%% file: 0.main.tex
\documentclass[sigconf]{acmart}
\usepackage{caption}
\usepackage{courier}
\usepackage{multirow}
\usepackage{wrapfig}
\usepackage{booktabs} % for professional tables
\usepackage{makecell}
\usepackage{graphicx}
\usepackage{xcolor}
\definecolor{myorange}{RGB}{2, 142, 2}
\usepackage{adjustbox}

\usepackage{amsmath}

\usepackage{gensymb}
\usepackage{amsfonts}       % blackboard math symbols
\usepackage{enumitem}
\usepackage{multirow}
\usepackage[linesnumbered, lined, ruled]{algorithm2e}
\usepackage{subfigure}
\usepackage{tikz}
\allowdisplaybreaks

\AtBeginDocument{%
  \providecommand\BibTeX{{%
    \normalfont B\kern-0.5em{\scshape i\kern-0.25em b}\kern-0.8em\TeX}}}

\copyrightyear{2022}
\acmYear{2022}
\setcopyright{acmcopyright}\acmConference[WWW '22]{Proceedings of the ACM Web Conference 2022}{April 25--29, 2022}{Virtual Event, Lyon, France}
\acmBooktitle{Proceedings of the ACM Web Conference 2022 (WWW '22), April 25--29, 2022, Virtual Event, Lyon, France}
\acmPrice{15.00}
\acmDOI{10.1145/3485447.3512202}
\acmISBN{978-1-4503-9096-5/22/04}

\acmSubmissionID{639}

\newcommand{\wy}[1]{{\color{black}{#1}}}
\begin{document}

\title{Learning Robust Recommenders through Cross-Model Agreement}

\author{Yu Wang}

\affiliation{%
  \institution{University of Science and Technology of China}
  \country{}
%   \city{Hefei}
%   \country{China}
}
\email{wy2001@mail.ustc.edu.cn}

\author{Xin Xin}
\authornote{Xin Xin is the corresponding author.}
\affiliation{%
  \institution{Shandong University}
  \country{}}
\email{xinxin@sdu.edu.cn}

\author{Zaiqiao Meng}
\affiliation{%
  \institution{Cambridge University}
  \country{}
}
\email{zaiqiao.meng@gmail.com}

\author{Joemon M Jose}
\affiliation{%
 \institution{University of Glasgow}
 \country{}}
\email{Joemon.Jose@glasgow.ac.uk}

\author{Fuli Feng}
\affiliation{%
 \institution{University of Science of Technology of China}
 \country{}}
\email{fulifeng93@gmail.com}

\author{Xiangnan He}
\affiliation{%
 \institution{University of Science and Technology of China}
 \country{}}
\email{xiangnanhe@gmail.com}

\begin{abstract}
Learning from implicit feedback is one of the most common cases in the application of recommender systems. Generally speaking, interacted examples are considered as positive while negative examples are sampled from uninteracted ones. However, noisy examples are prevalent in real-world implicit feedback. A \emph{noisy positive} example could be interacted but it actually leads to negative user preference. A \emph{noisy negative} example which is uninteracted because of user unawareness could also denote potential positive user preference. Conventional training methods overlook these noisy examples, leading to sub-optimal recommendations.  

In this work, we propose a general framework to learn robust recommenders from implicit feedback. Through an empirical study, we find that different models make relatively similar predictions on clean examples which denote the real user preference, while the predictions on noisy examples vary much more across different models. Motivated by this observation, we propose \em denoising with cross-model agreement \em (DeCA) which minimizes the KL-divergence between the real user preference distributions parameterized by two recommendation models while maximizing the likelihood of data observation. We instantiate DeCA on four representative recommendation models, empirically demonstrating its superiority over normal training and existing denoising methods. Codes are available at \url{https://github.com/wangyu-ustc/DeCA}.                                                                                                                                                                                                                                                                                                                                                                                                                            
\end{abstract}

\begin{CCSXML}
<ccs2012>
<concept>
<concept_id>10002950.10003648</concept_id>
<concept_desc>Mathematics of computing~Probability and statistics</concept_desc>
<concept_significance>500</concept_significance>
</concept>
<concept>
<concept_id>10002951.10003317.10003347.10003350</concept_id>
<concept_desc>Information systems~Recommender systems</concept_desc>
<concept_significance>500</concept_significance>
</concept>
</ccs2012>
\end{CCSXML}

\ccsdesc[500]{Mathematics of computing~Probability and statistics}
% \ccsdesc[300]{Computing methodologies~Machine learning}
\ccsdesc[500]{Information systems~Recommender systems}

\keywords{Recommender System, Denoised Recommendation, Robust Learning, Implicit Feedback.}


\maketitle

\input{1.introduction}
\input{2.motivation}
\input{3.method}

\input{4.experiments}

\input{5.related_work}
\input{6.conclusion}
% \clearpage
\begin{acks}
This work is supported by the National Key Research and Development Program of China (2020AAA0106000) and the National Natural Science Foundation of China (U19A2079, 62121002).
\end{acks}

\clearpage

\bibliographystyle{ACM-Reference-Format}
\bibliography{sample-base}
\clearpage

\appendix
\section{Supplement}
\subsection{Pseudo-code for DeCA and DeCA(p)} % (fold)
%Learning algorithms for DPI and DVAE are shown in Algorithm \ref{alg:DPI} and Algorithm \ref{alg:DVAE} respectively.
\vspace{-0.3cm}
\label{sec:algorithm_for_dpi_and_dvae}
\begin{algorithm}
\small
  % \SetKwInOut{Input}{Input}\SetKwInOut{Output}{Output}
  \KwIn{Corrupted data $\tilde{\textbf{R}}$, learning rate $\beta$, epochs $T$, hyper-parameter $\alpha, C_1, C_2$, regularization weight $\lambda$, target recommender $f$, auxiliary model $g$, $h$, $h'$}
  \KwOut{Parameters $\theta, \mu, \phi, \psi$ for $f,g,h,h'$, correspondingly}
  Initialize all parameters\;
  $count$ $\leftarrow$ 0\; 
  \While{Not early stop and $epoch < T$}{
    Draw a minibatch of $(u,i_+)$ from $\{(u,i)| \tilde{r}_{ui} = 1\}$\;
    Draw  $(u,i_-)$ from $\{(u,i)| \tilde{r}_{ui} = 0\}$\ for each $(u,i_+)$\;
    % Sample one negative item for each user in the minibatch and constitute $k$ minibatches of $(u,i_-)$ from $\{(u,i)| \tilde{r}_{ui} = 0\}$\;
    \uIf{$count \% 2 == 0$}{
      Compute $\mathcal{L}_{DeCA}$ according to Eq.(\ref{DPI-loss}) and Eq.(\ref{eq:likelihood-dfp})\;
    }
    \uElse{
      Compute $\mathcal{L}_{DeCA}$ according to Eq.(\ref{DPI-loss}) and Eq.(\ref{eq:likelihood-dfn})\;
    }
    Add regularization term: $\mathcal{L}_{DeCA} \leftarrow \mathcal{L}_{DeCA} + \lambda||\theta||^2$\;
    \For{each parameter $\Theta$ in $\{\theta, \mu, \phi, \psi\}$}{
      Compute $\partial \mathcal{L}_{DeCA} / \partial \Theta$ by back-propagation\;
      $\Theta \leftarrow \Theta - \beta \partial \mathcal{L}_{DeCA}/\partial \Theta$
    }
    $count$ $\leftarrow$ $count+1$\; 
  }
  \caption{Learning algorithm for DeCA}
  \label{alg:DPI}
\end{algorithm}
\vspace{-0.5cm}
\begin{algorithm}
\footnotesize
  \KwIn{Corrupted data $\tilde{\textbf{R}}$, learning rate $\beta$, epochs $T$, hyper-parameter $\alpha, C_1, C_2$, regularization weight $\lambda$, target recommender $f$, auxiliary model $h$, $h'$}
  \KwOut{Parameters $\theta, \phi, \psi$ for $f,h,h'$, correspondingly}
  Set random seed to $s_1$, initialize another copy of $\theta$ as $\theta'$\;
  \While{Not early stop and $epoch < T$}{
    Draw a minibatch of $(u,i_+)$ from $\{(u,i)| \tilde{r}_{ui} = 1\}$\;
    Draw  $(u,i_-)$ from $\{(u,i)| \tilde{r}_{ui} = 0\}$\ for each $(u,i_+)$\;
    Compute binary cross-entropy 
    $\mathcal{L}_{BCE}$ on $(u,i_+)$ and $(u,i_-)$ with $f_{\theta'}$\;
    Add regularization: $\mathcal{L}_{BCE} \leftarrow \mathcal{L}_{BCE} + \lambda||\theta'||^2$\;
    Compute $\partial \mathcal{L}_{BCE}/\partial \theta'$ by back-propagation\;
    $\theta' \leftarrow \theta' - \beta \partial \mathcal{L}/\partial \theta'$\;
  }
  Freeze $\theta'$, set random seed to $s_2$ and initialize $\theta, \phi, \psi$\; 
  $count$ $\leftarrow$ 0\; 
  \While{Not early stop and $epoch < T$}{
    Draw $(u,i_+)$ and $(u,i_-)$ similarly with line3-4\;
    \uIf{$count \% 2 == 0$}{
      Compute $\mathcal{L}_{DeCA(p)}$ according to Eq.(\ref{DVAE-loss}) and Eq.(\ref{eq:likelihood-dfp})\;
    }
    \uElse{
      Compute $\mathcal{L}_{DeCA(p)}$ according to Eq.(\ref{DVAE-loss}) and Eq.(\ref{eq:likelihood-dfn})\;
    }
    Add regularization term: $\mathcal{L}_{DeCA(p)} \leftarrow \mathcal{L}_{DeCA(p)} + \lambda||\theta||^2$\;
    \For{each parameter $\Theta$ in $\{\theta, \phi, \psi\}$}{
      Compute $\partial \mathcal{L}_{DeCA(p)} / \partial \Theta$ by back-propagation\;
      $\Theta \leftarrow \Theta - \beta \partial \mathcal{L}_{DeCA(p)}/\partial \Theta$\;
    }
    $count$ $\leftarrow$ $count+1$\; 
  }
  \caption{Learning algorithm for DeCA(p)}
  \label{alg:DVAE}
\end{algorithm}
% section algorithm_for_dpi_and_dvae (end)

\subsection{Mathematical Formulations}
\label{ssub:mathematical_formulation}
The detailed formulation of the term $E_{\textbf{R}\sim P_f}[\log P(\tilde{\textbf{R}}|\textbf{R})]$ in $\mathcal{L}_{DeCA}$ is shown as: 
\begin{align}
\small
  &E_{\textbf{R}\sim P_f}[\log P(\tilde{\textbf{R}}|\textbf{R})] = \sum_{(u,i)} E_{r_{ui}\sim P_f} [\log P(\tilde{r}_{ui}|r_{ui})] \nonumber \\
  &=\sum_{(u,i)|\tilde{r}_{ui}=1}  \left\{\begin{array}{ll}
    \log P(\tilde{r}_{ui}=1|r_{ui}=1) \cdot P_f(r_{ui}=1) \\+
     \log P(\tilde{r}_{ui}=1|r_{ui}=0)\cdot P_f(r_{ui}=0) \end{array}\right. \nonumber\\
     &+\sum_{(u,i)|\tilde{r}_{ui}=0} \left\{\begin{array}{ll}
    \log P(\tilde{r}_{ui}=0|r_{ui}=1) \cdot P_f(r_{ui}=1) \\
    + \log P(\tilde{r}_{ui}=0|r_{ui}=0)\cdot P_f(r_{ui}=0)
  \end{array}\right. \nonumber \\
  &=\hspace{-0.3cm}\sum_{(u,i)|\tilde{r}_{ui}=1}\hspace{-0.3cm}
    \log h'_\psi(u,i) \cdot f_{\theta}(u,i) + \log h_\phi(u,i) \cdot (1-f_\theta(u,i))\nonumber \\
  &+\hspace{-0.3cm}\sum_{(u,i)|\tilde{r}_{ui}=0}\hspace{-0.3cm}
  \log(1 - h'_\psi(u,i))\cdot f_{\theta}(u,i) + \log(1 - h_\phi(u,i) ) \cdot (1-f_\theta(u,i)).
  \label{eq:likelihood}
\end{align}
The KL divergence term $D[P_g||P_f]$ can be calculated as
\begin{equation*}
  D[P_g||P_f] = g_\mu(u,i) \cdot \log \frac{g_\mu(u,i)}{f_\theta(u,i)} + (1-g_\mu(u,i)) \cdot \log \frac{1 - g_\mu(u,i)}{1-f_\theta(u,i)}.
\end{equation*}
$D[P_f||P_g]$ is computed similarly with $D[P_g||P_f]$.

\subsection{Additional Discussion} % (fold)
In this subsection, we provide a brief discussion to illustrate the relationship between our methods and other related methods.  
\subsubsection{Relationship with Variational Auto-Encoder}
We can see that Eq.(\ref{VAE_lower_bound}) is exactly the objective function of a variational auto-encoder (VAE) \cite{doersch2016vaetutorial}.
Specifically, the real user preference $\textbf{R}$ is the latent variable. $P_f(\textbf{R}|\tilde{\textbf{R}})$ which is parameterized by our target model $f_\theta$ maps the corrupted data $\tilde{\textbf{R}}$ to the latent variables $\textbf{R}$ and thus can be seen as the encoder. The likelihood $P(\tilde{\textbf{R}}|\textbf{R})$ describes the distribution of corrupted data $\tilde{\textbf{R}}$ given the real user preference $\textbf{R}$ and acts as the decoder. Finally, $P(\mathbf{R})$ is the fixed prior distribution. However, we claim that the proposed DeCA(p) is substantially different from VAE. For example, DeCA(p) does not utilize the re-parameterization of VAE. VAE is just an interpretation of DeCA(p). Besides, although there are some methods utilizing VAE for recommendation ~\cite{DBLP:conf/www/LiangKHJ18,DBLP:conf/wsdm/ShenbinATMN20}, few of them are designed for recommendation denoising. Finally, existing VAE-based recommendation methods are actually recommendation models while the proposed DeCA(p) acts a \emph{model-agnostic training framework} which is then instantiated with all kinds of downstream models. They serve on different levels.  
% \subsubsection{Relationship with Ensemble Approaches}
% The proposed methods utilize predictions from different models for robust recommendations. However, we claim that there are obvious differences between our proposed methods and ensemble approaches. In ensemble methods, each model serves the same task and the forward pass needs to be carried out on every model to ensemble during the inference stage. However, in our methods, different models function differently and we only need the target model for inference. It means that our model is more time-efficient in the inference stage compared with ensemble approaches. 

\begin{table*}[h!]
\centering
\caption{Overall performance comparison. The highest scores are in Boldface. R is short for Recall and N is short for NDCG. The results with improvements over the best baseline larger than 5\% are marked with $*$.}
%\vspace{-0.3cm}
\label{tab:Overall_performance2}
\resizebox{\textwidth}{!}{%
\begin{tabular}{c|cccc|cccc}
\toprule
 \multirow{2}{*}{\textbf{Electronics}} & R@10 & R@50 &N@10 &N@50 & R@10 & R@50 &N@10 &N@50 \\
\cmidrule(lr){2-5}\cmidrule{6-9} 
 & \multicolumn{4}{c|}{GMF} & \multicolumn{4}{c}{NeuMF} \\ 
    \hline
Normal & 0.023$\pm$0.000 & 0.063$\pm$0.001 & 0.013$\pm$0.000 & 0.021$\pm$0.000 & 0.075$\pm$0.000 & 0.191$\pm$0.002 & 0.048$\pm$0.000 & 0.072$\pm$0.001 \\
WBPR & 0.026$\pm$0.001 & 0.072$\pm$0.001 & 0.014$\pm$0.000 & 0.024$\pm$0.000 & \textbf{0.077}$\pm$0.000 & \textbf{0.201}$\pm$0.001 & \textbf{0.048}$\pm$0.000 & 0.074$\pm$0.000 \\
T-CE & 0.024$\pm$0.000 & 0.063$\pm$0.001 & 0.013$\pm$0.000 & 0.022$\pm$0.000 & 0.071$\pm$0.004 & 0.189$\pm$0.008 & 0.045$\pm$0.002 & 0.071$\pm$0.003 \\
Ensemble & 0.014$\pm$0.001 & 0.034$\pm$0.001 & 0.010$\pm$0.000 & 0.016$\pm$0.000 &  0.051$\pm$0.000 & 0.108$\pm$0.001 & 0.040$\pm$0.000 & 0.055$\pm$0.000 \\
DeCA & \textbf{0.045$\pm$0.001$^*$} & 0.104$\pm$0.000 & 0.023$\pm$0.000 & 0.037$\pm$0.000 & 0.071$\pm$0.001 & 0.191$\pm$0.001 & 0.044$\pm$0.002 & 0.069$\pm$0.001 \\
DeCA(p) & 0.042$\pm$0.001 & \textbf{0.110$\pm$0.000$^*$} & \textbf{0.024$\pm$0.000$^*$} & \textbf{0.038$\pm$0.000$^*$} & 0.074$\pm$0.002 & 0.200$\pm$0.001 & 0.047$\pm$0.000 & \textbf{0.074}$\pm$0.000 \\
\hline
& \multicolumn{4}{c|}{CDAE} & \multicolumn{4}{c}{LightGCN} \\ 
\hline
Normal & 0.074$\pm$0.001 & 0.197$\pm$0.001 & 0.047$\pm$0.000 & 0.073$\pm$0.000 & 0.024$\pm$0.000 & 0.081$\pm$0.001 & 0.016$\pm$0.000 & 0.030$\pm$0.000 \\
WBPR & 0.077$\pm$0.000 & 0.190$\pm$0.000 & 0.048$\pm$0.000 & 0.072$\pm$0.000 & 0.022$\pm$0.000 & 0.076$\pm$0.001 & 0.015$\pm$0.000 & 0.028$\pm$0.000 \\
T-CE & 0.067$\pm$0.002 & 0.187$\pm$0.010 & 0.044$\pm$0.001 & 0.067$\pm$0.003 & 0.024$\pm$0.001 & 0.086$\pm$0.001 & 0.017$\pm$0.000 & 0.031$\pm$0.000 \\
Ensemble & 0.043$\pm$0.000 & 0.110$\pm$0.001 & 0.036$\pm$0.000 & 0.056$\pm$0.000 & 0.018$\pm$0.000 & 0.049$\pm$0.001 & 0.014$\pm$0.000 & 0.023$\pm$0.000 \\
DeCA & \textbf{0.077}$\pm$0.000 & 0.201$\pm$0.001 & 0.048$\pm$0.000 & 0.075$\pm$0.000 & 0.025$\pm$0.000 & 0.103$\pm$0.001 & 0.018$\pm$0.000 & 0.036$\pm$0.000 \\
DeCA(p) & 0.077$\pm$0.000 & \textbf{0.202}$\pm$0.001 & \textbf{0.048}$\pm$0.000 & \textbf{0.075}$\pm$0.000 & \textbf{0.026}$\pm$0.001 & \textbf{0.104$\pm$0.001$^*$} & \textbf{0.019}$\pm$0.000 & \textbf{0.037$\pm$0.000$^*$} \\
\bottomrule
\end{tabular}}
% \vspace{-5pt}
\end{table*}
\begin{table*}[h!]
    \centering
    \caption{Effect of model selection on MovieLens, Adressa and Electronics}
    \begin{tabular}{c|c|cccc|cccc|cccc}
    \toprule
    & & \multicolumn{4}{c|}{MovieLens} & \multicolumn{4}{c|}{Addressa} & \multicolumn{4}{c}{Electronics} \\
        Method & $h$ and $h'$ & R@5 & R@20 & N@5 & N@20 &  R@5 & R@20 & N@5 & N@20 & R@10 & R@50 & N@10 & N@50 \\
    \midrule
    \multicolumn{2}{c|}{Normal} & 0.0214 & 0.0953 &  0.0352 & 0.0593 & 0.1163 & 0.2089 &  0.0801 & 0.1125 & 0.0227 & 0.0633 & 0.0128 & 0.0214 \\
    \midrule
    \multirow{3}{*}{DeCA}& MF& 0.0217 & 0.1089 & 0.0382 & 0.0643 & 0.1255 & 0.2253 & 0.0907 & 0.1250 & 0.0451 & 0.1040 & 0.0234 & 0.0365 \\
    & GMF & 0.0186 & 0.1054 & 0.0351 & 0.0634 & 0.1239 & 0.2259 & 0.0908 & 0.1260 & 0.0447 & 0.1047 & 0.0235 & 0.0368 \\
    & NeuMF&0.0225 & 0.1077 & 0.0349 & 0.0621 & 0.1268 & 0.2270 & 0.0915 & 0.1260 & 0.0447 & 0.1042 & 0.0233 & 0.0365 \\
\hline
\multirow{3}{*}{DeCA(p)}& MF& 0.0269 & 0.0987 & 0.0538 & 0.0658 & 0.1233 & 0.2203 & 0.0932 & 0.1270 & 0.0411 & 0.1102 & 0.0222 & 0.0372 \\
& GMF & 0.0262 & 0.0980 & 0.0536 & 0.0646 & 0.1221 & 0.2182 & 0.0912 & 0.1245 & 0.0428 & 0.1125 & 0.0235 & 0.0386 \\
& NeuMF& 0.0255 & 0.0963 & 0.0539 & 0.0653 & 0.1216 & 0.2178 & 0.0906 & 0.1240 & 0.0429 & 0.1115 & 0.0237 & 0.0386  \\
\bottomrule
    \end{tabular}
    \label{tab:effect of model selection}
\end{table*}
\begin{figure*}[h]
\centering     %%% not \center
\subfigure[DeCA - MovieLens]{\label{fig:DPI_movielens_hyer}\includegraphics[width=0.15\linewidth]{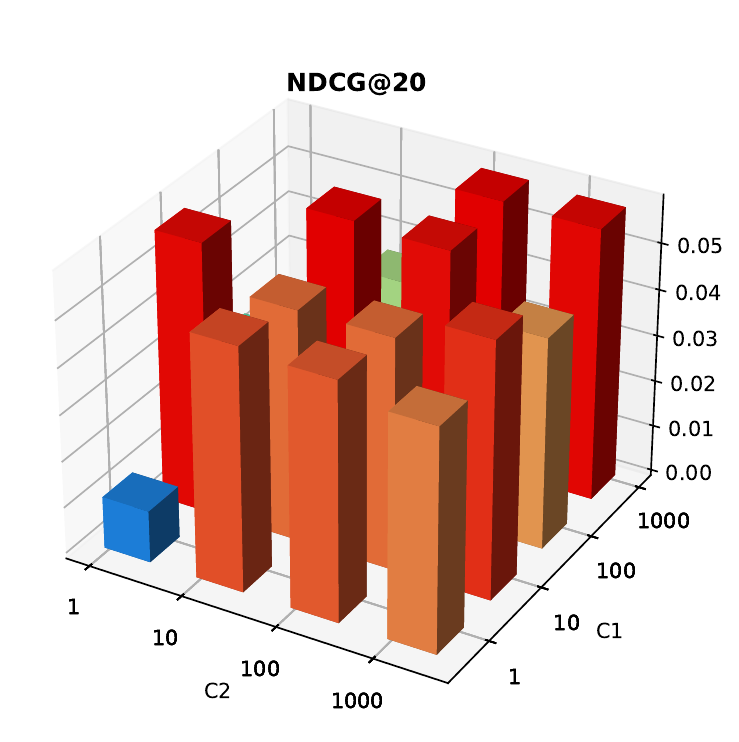}}
\subfigure[DeCA(p) - MovieLens]{\label{fig:DVAE_movielens_hyer}\includegraphics[width=0.15\linewidth]{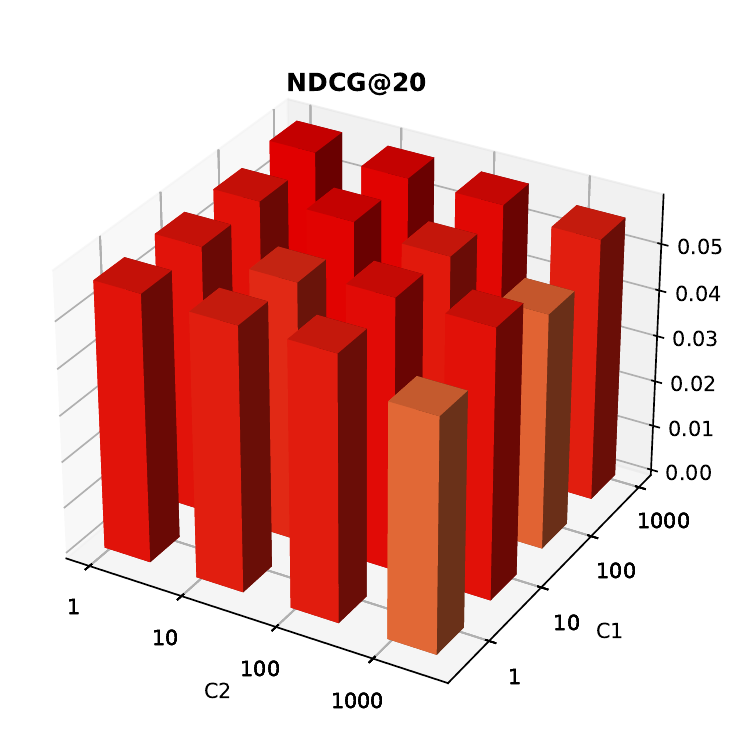}}
\subfigure[DeCA - Modcloth]{\label{fig:DPI_GMF_modcloth_hyper}\includegraphics[width=0.165\linewidth]{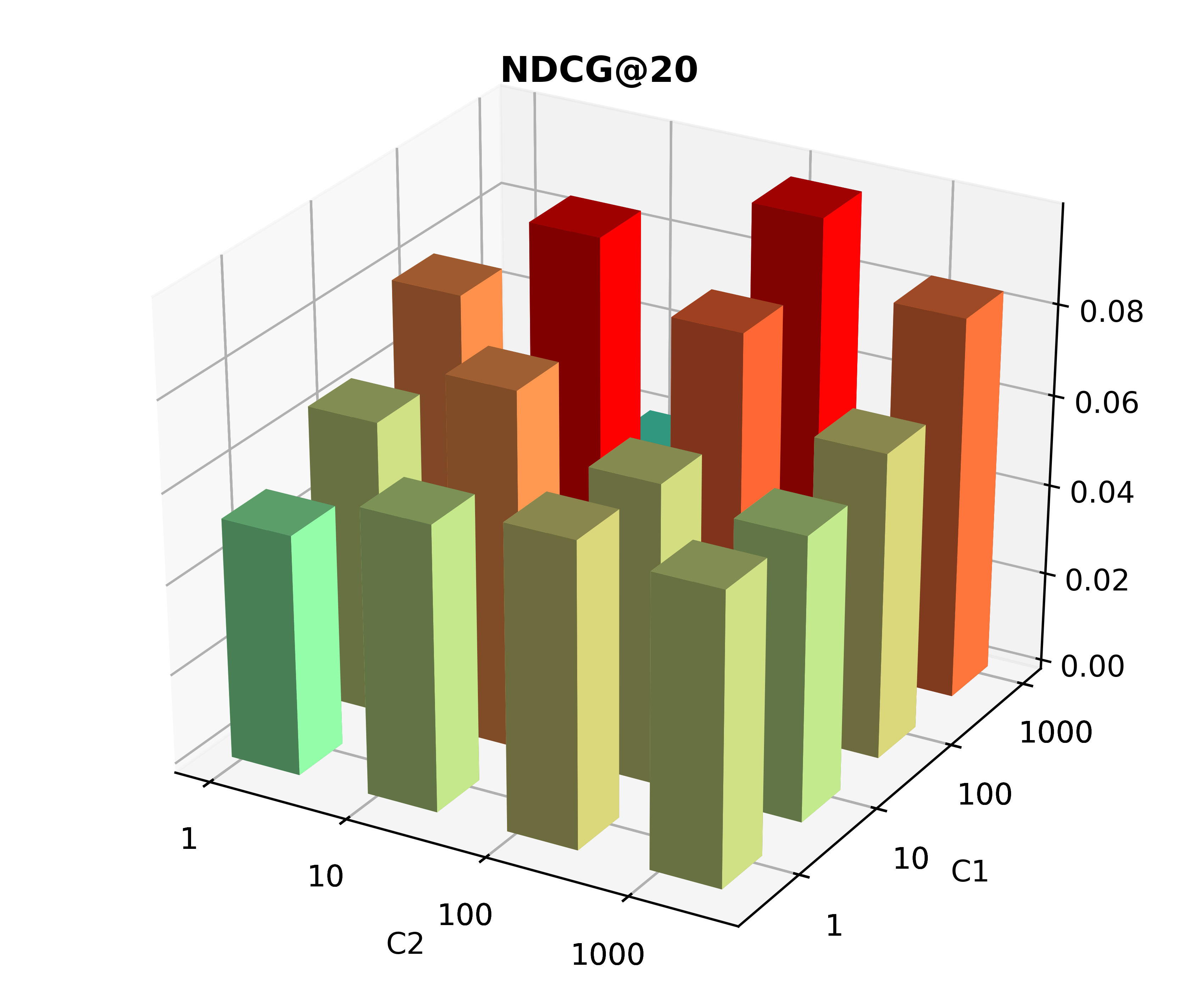}}
\subfigure[DeCA(p) - Modcloth]{\label{fig:DVAE_GMF_modcloth_hyper}\includegraphics[width=0.165\linewidth]{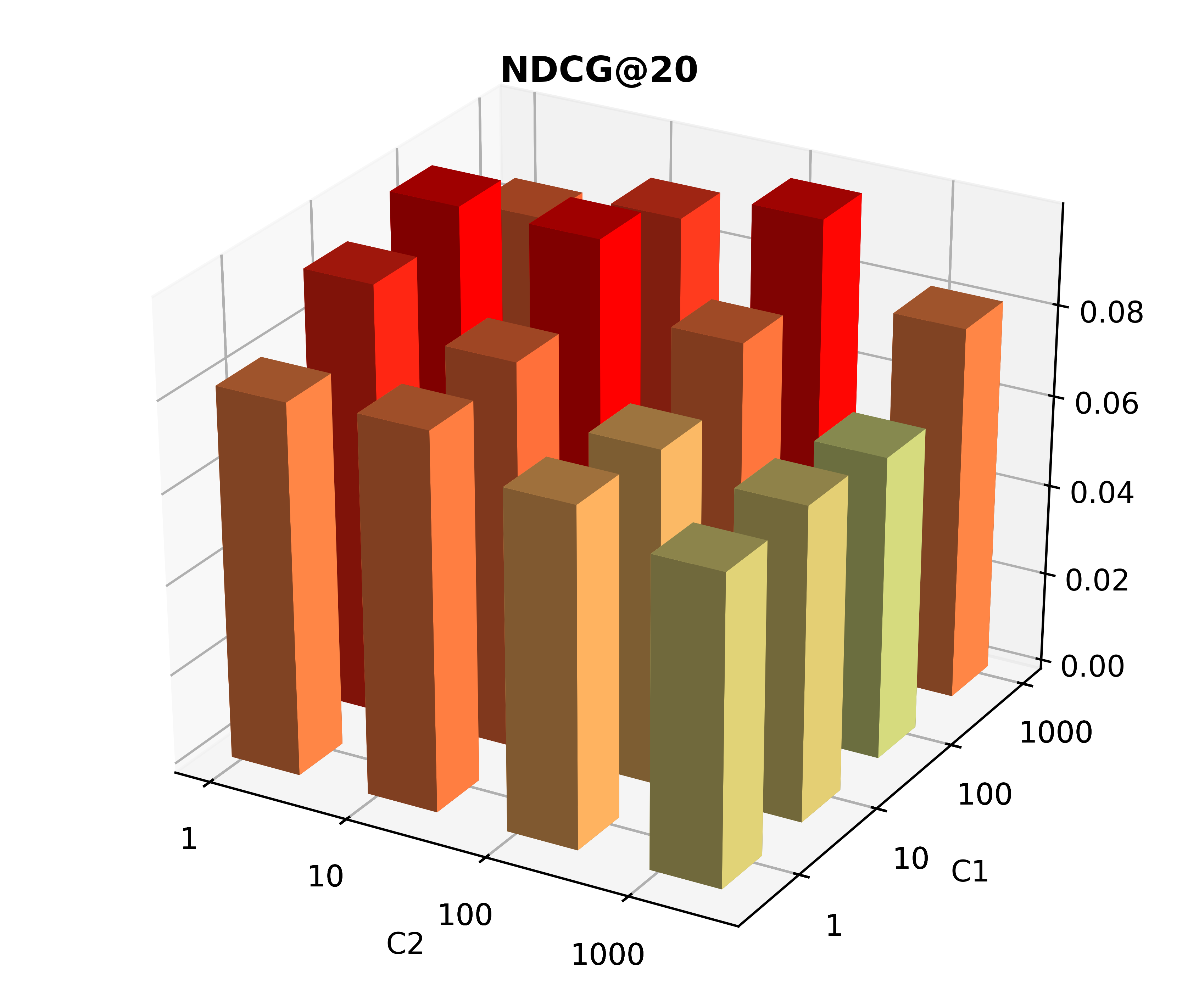}}
\subfigure[DeCA - Electronics]{\label{fig:DPI_GMF_electronics_hyper}\includegraphics[width=0.165\linewidth]{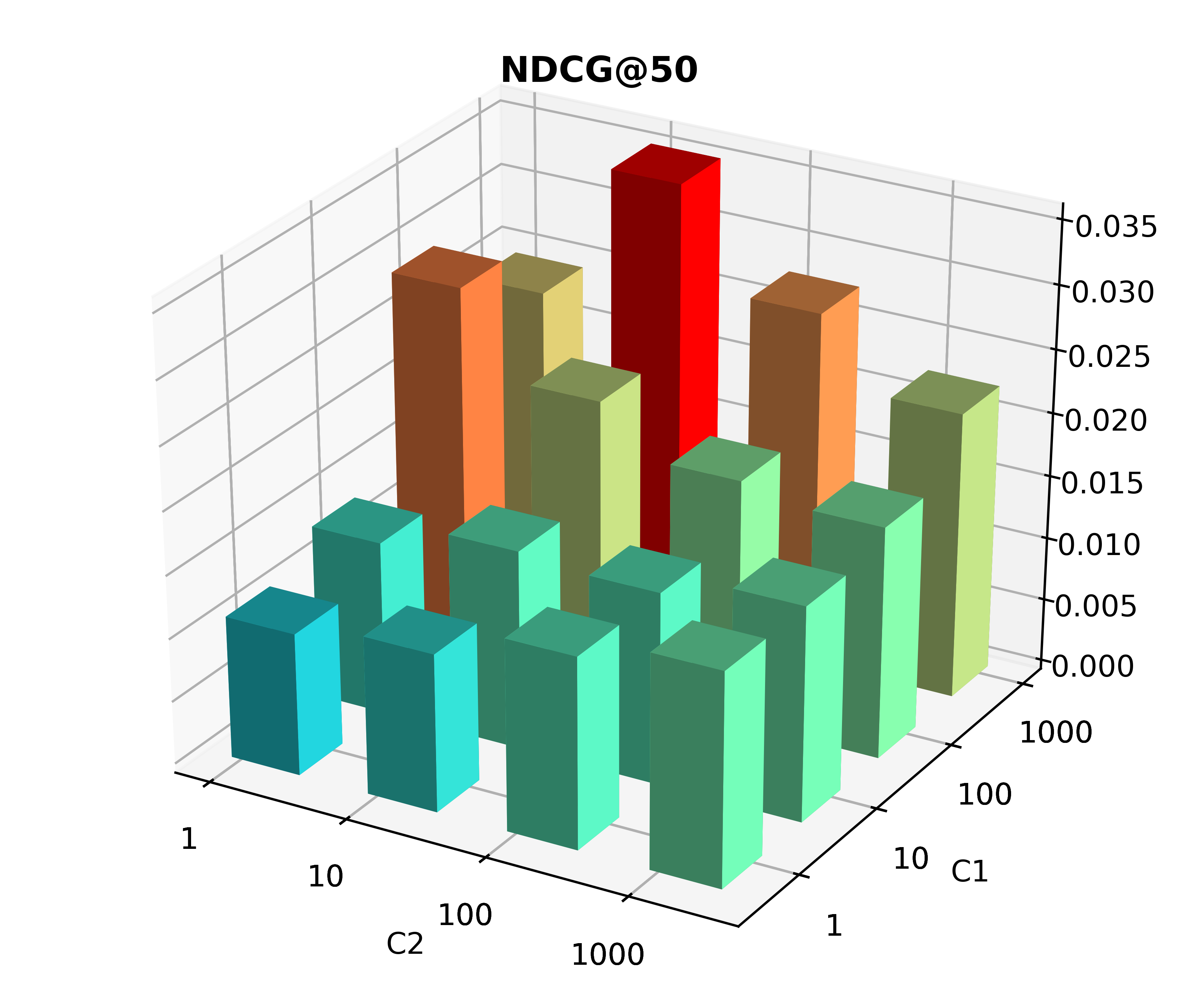}}
\subfigure[DeCA(p) - Electronics]{\label{fig:DVAE_GMF_electronics_hyper}\includegraphics[width=0.165\linewidth]{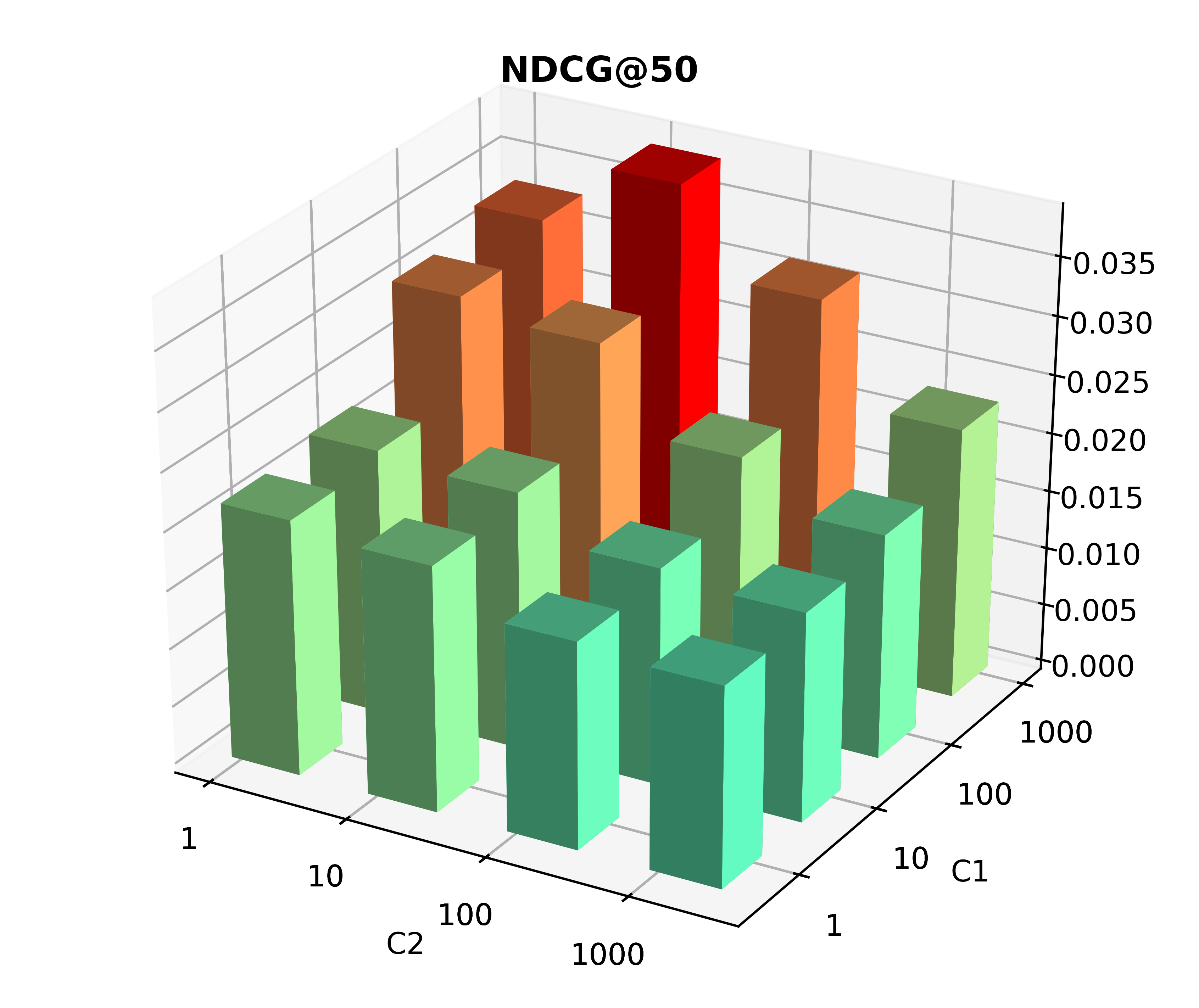}}
\caption{Hyperparameter study on Movielens, Modcloth and Electronics.}
\label{fig:hyper_parameter_study_for_modcloth_and_electronics}
\end{figure*}

\wy{
\subsubsection{Analysis of the Convergence}
During training, if the parameters $C_1$ and $C_2$ are large enough, the process will act similarly to the normal training, since (1) the gradients from the fixed part will dominate the update of the model when $C_1$ and $C_2$ are large. (2) With large $C_1$ and $C_2$, we will demonstrate that the objective has the same target as cross-entropy loss as follows. 

If we simply replace $-\log(1-h'_\psi(u, i))$ and $-\log(1-h_\phi(u, i))$ with $C_1$ and $C_2$ in the term $E[\log P(\tilde{\textbf{R}}|\textbf{R})]$, we could obtain the following expansion: 
\begin{equation}\label{eq:expansion}
    E[\log P(\tilde{\textbf{R}}|\textbf{R})] = - \sum_{(u,i)|\tilde{r}_{ui} = 0} C_1 \cdot f_\theta(u,i) - \sum_{(u,i)|\tilde{r}_{ui} = 1} C_2 \cdot (1 - f_\theta(u,i))
\end{equation}
Thus maximizing Eq.(\ref{eq:expansion}) will optimize $f_\theta(u,i)$ to fit the observational data. The loss for the dataset is similar to cross-entropy loss, and it will converge. 

\subsubsection{Analysis of the Run Time Complexity}
For DeCA, the main additional complexity comes from the forward and backward of the MF model $g_\mu$. However, since MF is the simplest model, the time complexity will be much smaller than twice the complexity of normal training. 

For DeCA(p), the total time complexity will be almost exactly twice of normal training, since  the complexity of the second time is almost the same as the first time, i.e., normal training.
}

\subsection{Supplementary Experimental Results}
\subsubsection{Performance Comparison on Electronics}
The performance comparison between models and baselines on the Electronics dataset is shown in Table \ref{tab:Overall_performance2}.
\subsubsection{Supplementary Results for Hyperparameter Study}
Figure \ref{fig:hyper_parameter_study_for_modcloth_and_electronics} shows the supplementary results for hyperparameter study on MovieLens, Modcloth and Electronics.

\subsubsection{Supplementary Results for Model Selection}
Table \ref{tab:effect of model selection} shows the supplementary results for the selection of $h$ and $h'$ on MovieLens, Adressa and Electronics.
\clearpage
\end{document}

%% file: 1.introduction.tex
% \vspace{-5pt}
\section{Introduction}
\begin{figure*}
\centering     %%% not \center
\subfigure[Normal training]{\label{fig:nomral}\includegraphics[width=0.245\linewidth]{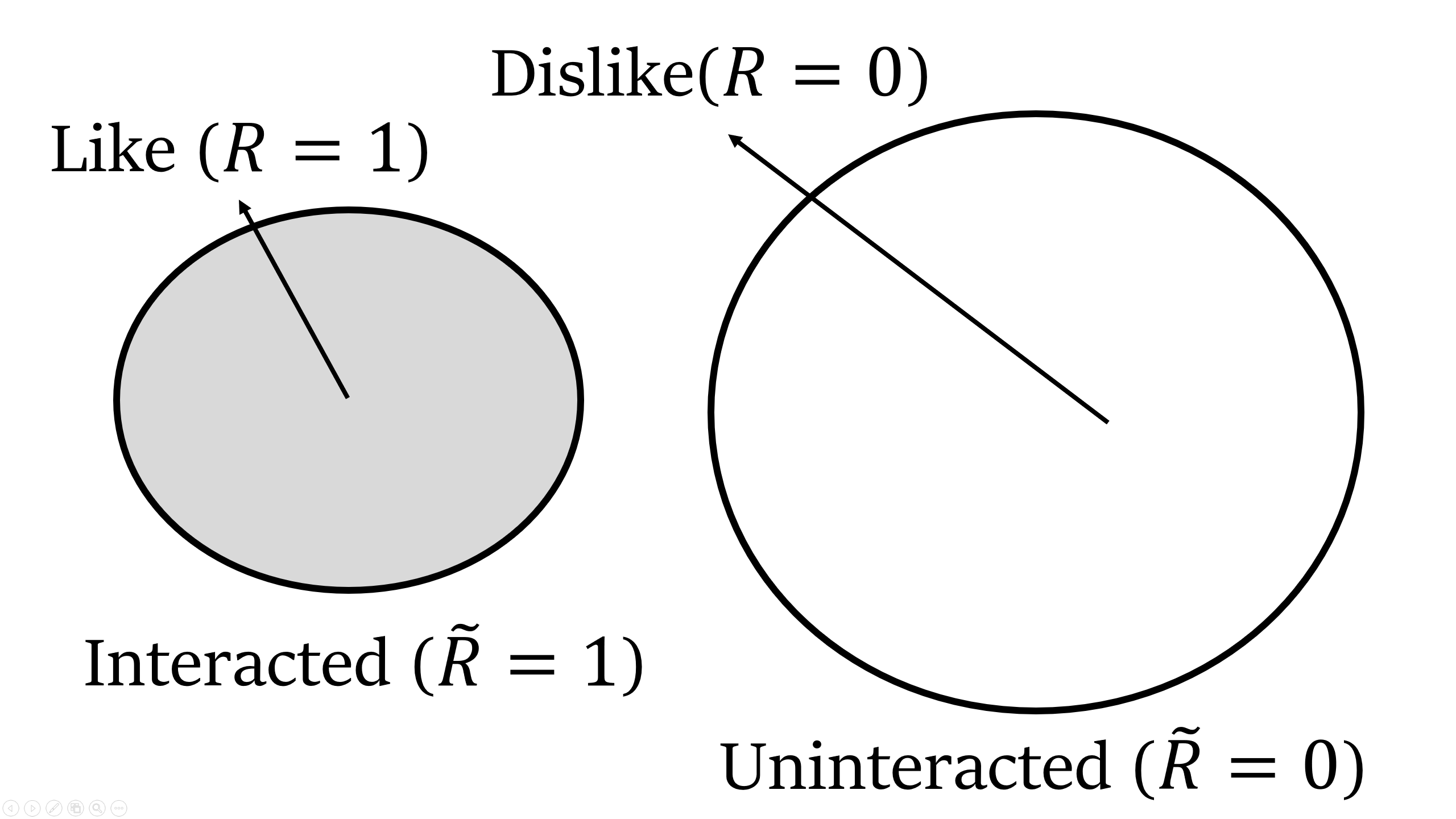}}
\subfigure[Real-world scenario]{\label{fig:FB}\includegraphics[width=0.245\linewidth]{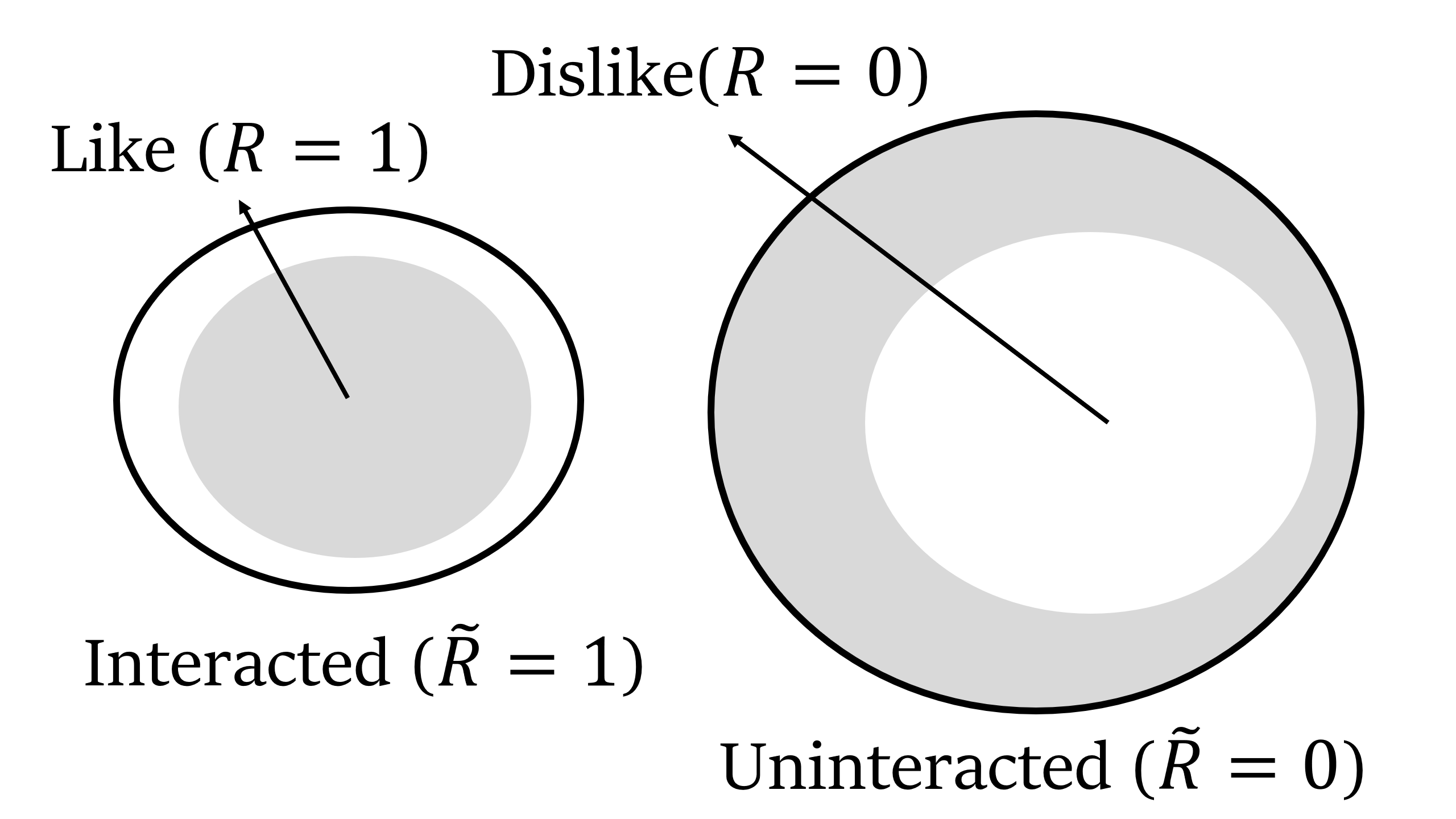}}
\subfigure[Denoising Positive ]{\label{fig:FN}\includegraphics[width=0.245\linewidth]{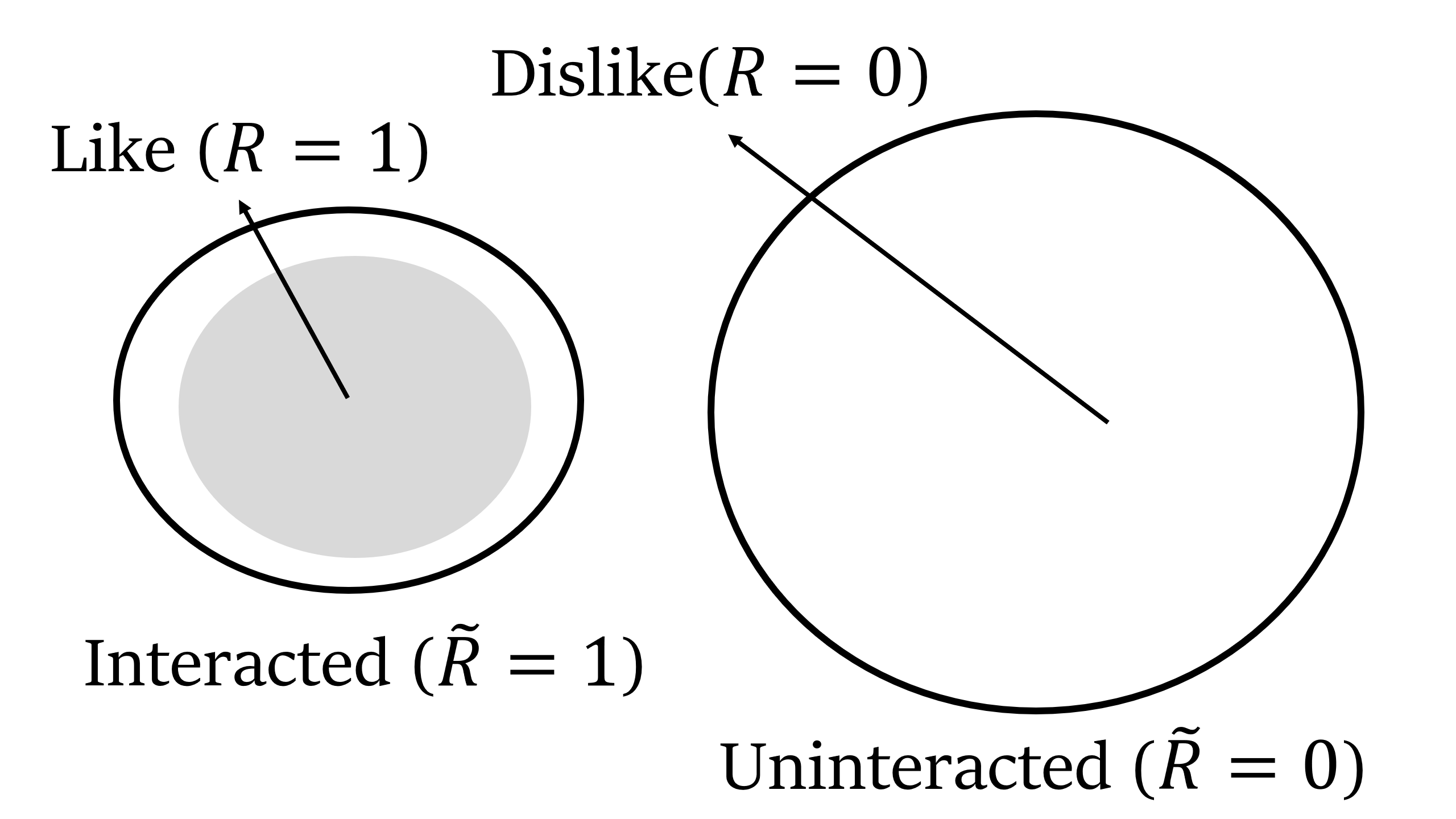}}
\subfigure[Denoising Negative ]{\label{fig:FP}\includegraphics[width=0.245\linewidth]{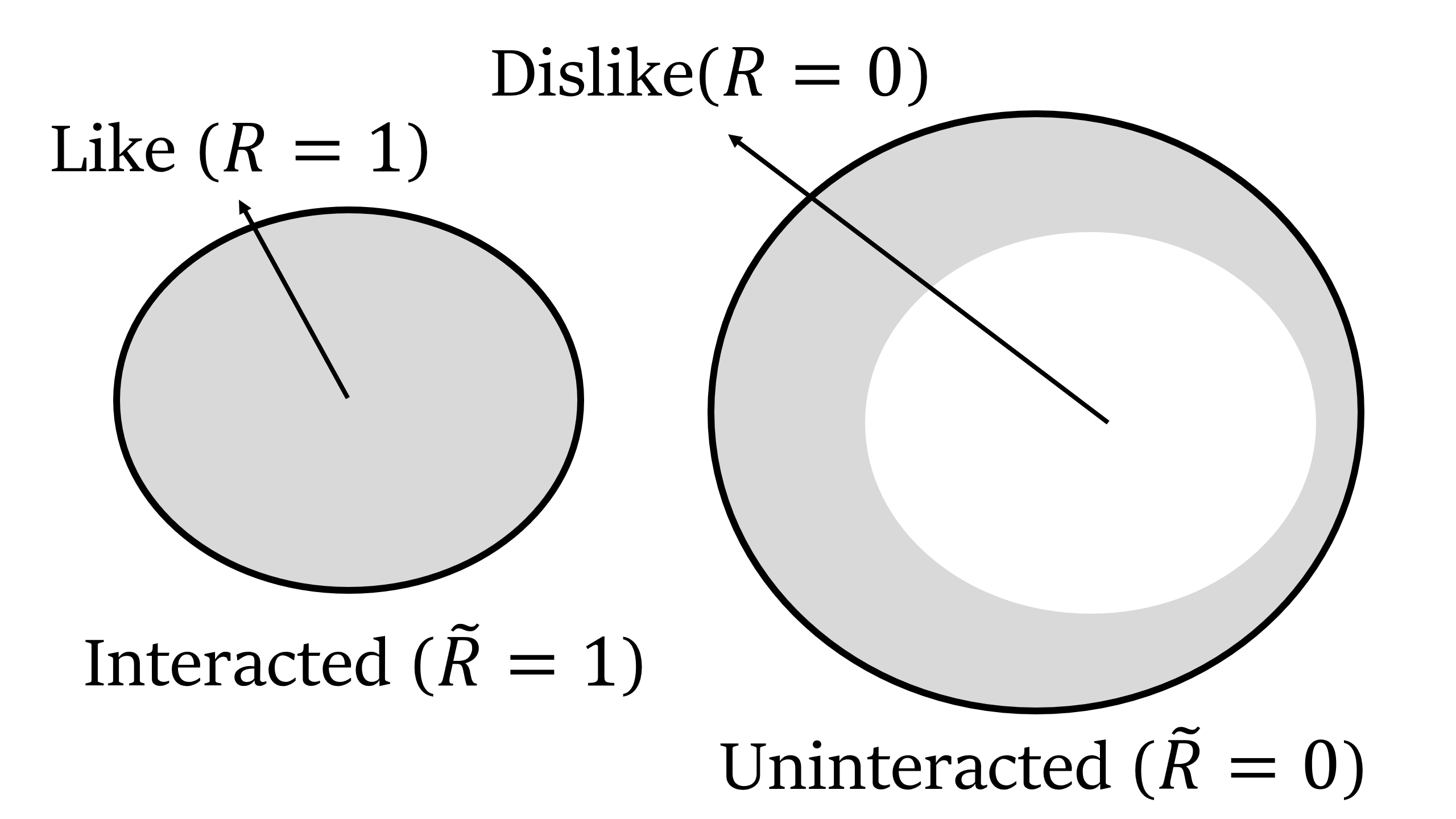}}
\vspace{-0.5cm}
\caption{Different assumptions in recommendation training on implicit feedback. Gray denotes positive user preference (i.e., `like') while white denotes negative preference (i.e.,  `dislike'). In normal training, interacted instances are considered as positive examples and uninteracted ones are considered as negative examples, as shown in (a). 
However, in real-world applications, there are both  noisy-positive examples in interacted items and noisy-negative examples in uninteracted items, as shown in (b). (c) denotes only denoising positive examples \cite{wang2020denoising,lu2019effects}, while (d) only denoises negative examples \cite{yu2020sampler,ding2018improvedview}.}
\vspace{-0.4cm}
\end{figure*}
Recommender systems are widely used for information filtering, especially for various online services such as E-commerce \cite{lin2019cross}, multimedia platforms \cite{davidson2010youtube,van2013deepspotify} and social media \cite{chen2019efficient}.
Among existing systems, training using implicit feedback (e.g. view and click behaviors) is more prevalent than that using explicit ratings, since implicit feedback is much easier to be collected \cite{rendle2014bayesian,he2016fast}.
Specifically, the interacted user-item samples
are assumed to be positive examples while negative examples are sampled from missing 
interactions.
Under this assumption, the interacted examples are regarded to denote positive user preference (i.e., like) and thus are trained towards higher prediction scores. The sampled missing interactions are considered as negative user preferences (i.e., dislike) and are pushed down in the training procedure. Figure \ref{fig:nomral} illustrates this normal training assumption.

However, this assumption seldom holds for real-world applications since noisy examples are prevalent, as shown in Figure \ref{fig:FB}. For example, an interacted user-item pair may only be attributed to the popularity bias \cite{chen2020bias}, which could actually lead to negative user preference. These interactions can be considered as \emph{noisy-positive} examples \cite{lu2018between, hu2008collaborative, wen2019leveraging}. Besides, a missing interaction could also be attributed to user unawareness because the item is not exposed to this user.
Such \emph{noisy-negative} examples could have signaled potential positive user preferences. 
The normal training procedure overlooks these noisy examples, leading to the misunderstanding of the real user preference and sub-optimal recommendation \cite{chen2020bias,wen2019leveraging}.  

To denoise implicit feedback, some recent efforts have been done by using re-sampling methods \cite{yu2020sampler,gantner2012wbpr,ding2019samplerview,ding2018improvedview,DBLP:conf/sigir/WangYZGXWZZ17} or re-weighting methods \cite{wang2020denoising}.
Re-sampling methods focus on designing more effective samplers. For example, \cite{gantner2012wbpr} considers that the missing interactions of popular items are highly likely to be real negative examples. However, the performance of re-sampling methods depends heavily on the sampling distribution and suffers from high variance \cite{yuan2018fbgd}. \cite{wang2020denoising} proposed a re-weighting method that assigns lower weights or zeroes weights to high-loss samples since the noisy examples would have higher losses. However, \cite{shu2019meta} shows that hard yet clean examples also tend to have high losses. As a result, \cite{wang2020denoising} could encounter difficulties distinguishing between hard clean and noisy examples. Besides, some research focuses on utilizing auxiliary information to denoise implicit feedback \cite{lu2019effects,kim2014modeling,liu2010understanding,yi2014beyond}  but these kinds of methods need additional input.

In this work, we propose a general training framework to learn robust recommenders from implicit feedback without introducing external knowledge sources. The signal
for denoising comes from an insightful observation: different models tend to make relatively similar predictions for clean examples which represent the real user preference, while the predictions for noisy examples would vary much more among different models. (See figure \ref{fig:prediction_differences} for more details.)
To this end, we propose \emph{denoising with cross-model agreement} (DeCA) which minimizes the Kullback–Leibler (KL) divergence between the real user preference distributions parameterized by two recommendation models, and meanwhile, maximizes the likelihood of the data observation given the real user preference. DeCA can be considered as a framework that incorporates the weakly supervised signals from different model predictions to denoise the target recommender. 
The proposed DeCA acts as a robust learning framework and can be naturally incorporated with any recommendation model. 

To summarize, the main contributions of this work are as follows:
\begin{itemize}[leftmargin=*]
  \item We find that the different models tend to make more similar predictions for clean examples than for noisy ones. This observation provides new hints to devise denoising methods for implicit feedback recommendation.
  \item We propose the robust learning frameworks DeCA which utilizes the agreement across different model predictions as the denoising signal for implicit feedback and then infers the real user preference from corrupted binary data.
  \item We instantiate DeCA with four classical recommendation models and conduct extensive experiments on four datasets. The results demonstrate the effectiveness of our proposal.
\end{itemize}

%% file: 2.motivation.tex
\section{Motivation}
\label{sec:motivation}
According to the theory of robust learning, \cite{han2018co,jiang2018mentornet}, noisy examples tend to have relatively high loss values. However, the hard yet clean examples which fall near the classification boundary also have high losses \cite{shu2019meta}. These hard clean examples play an important role to boost the recommendation performance of implicit feedback \cite{yuan2016lambdafm,ding2019reinforced,wang2020reinforced}. Denoising methods based on loss values %\zmc{Cannot understand `based on loss values' here.} 
could encounter difficulties distinguishing between noisy examples and hard clean examples, leading to sub-optimal solutions. However, we find that another signal which can help to identify noisy examples can be obtained from the agreement across different model predictions.  

\begin{figure}
\vspace{-0.2cm}
\centering
\subfigure[Different models.]{\label{fig:Differences_in_models}\includegraphics[width=0.495\linewidth]{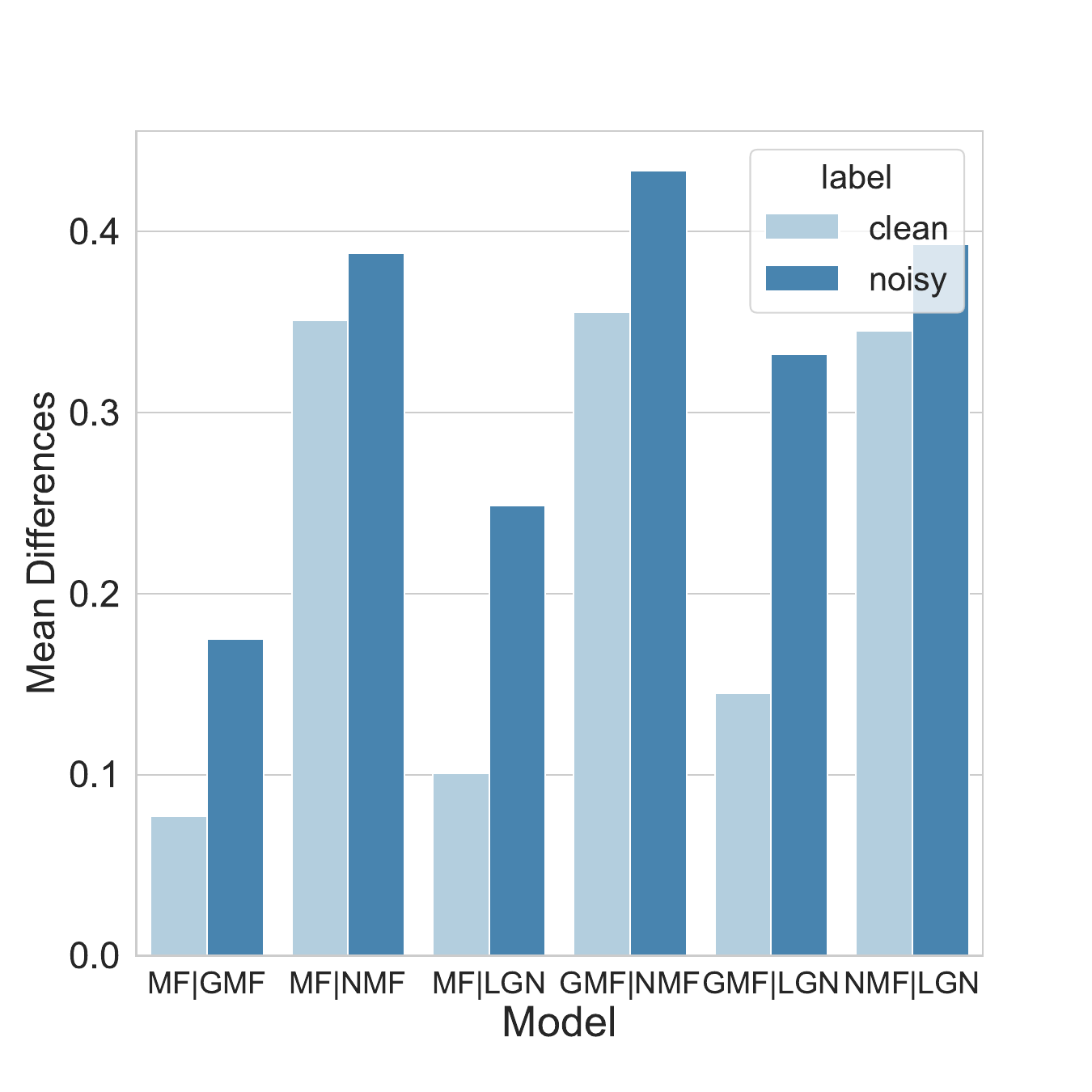}}
\subfigure[Different random seeds.]{\label{fig:Different_random_seeds}\includegraphics[width=0.495\linewidth]{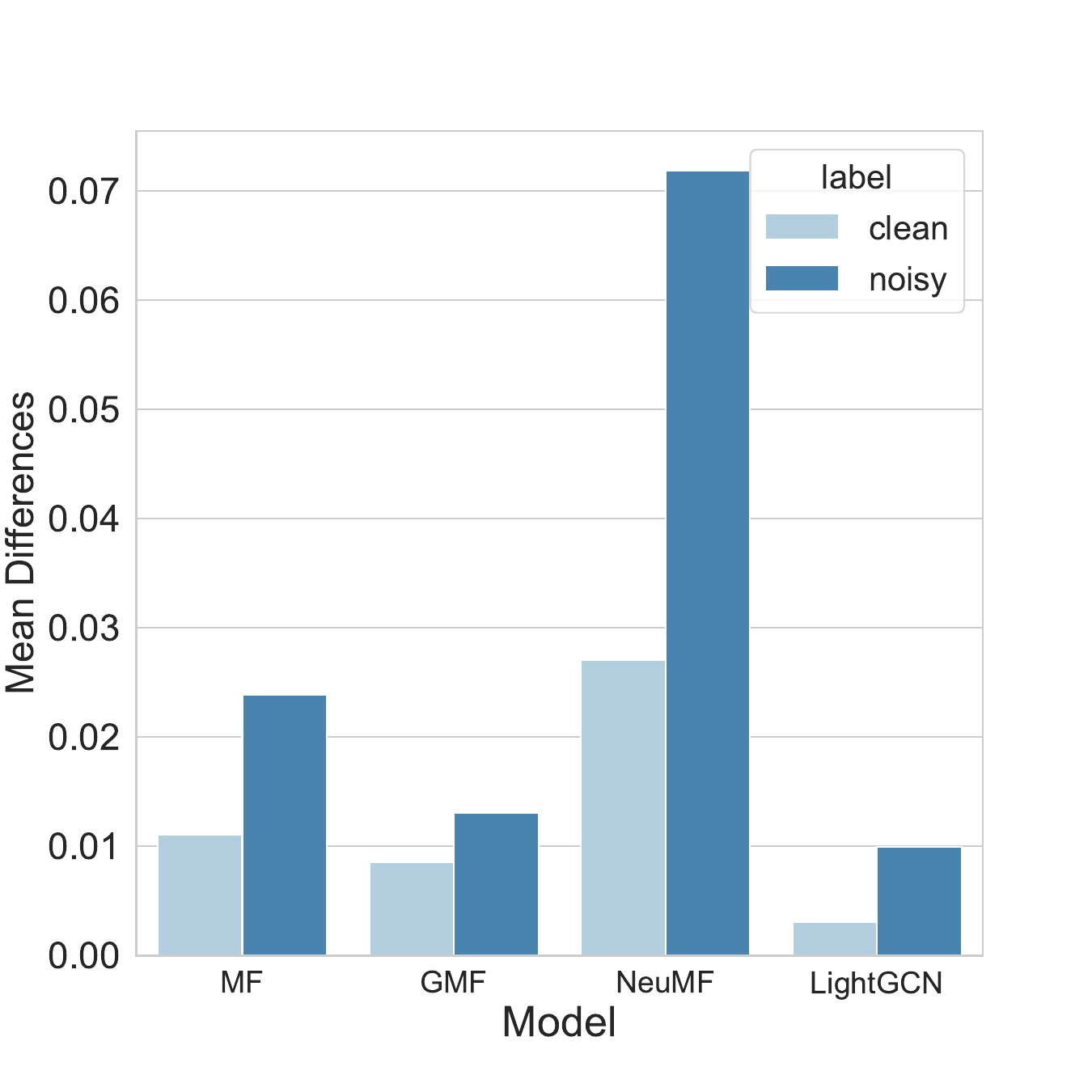}}
\vspace{-0.3cm}
\caption{Mean prediction differences on \textbf{clean} and \textbf{noisy} examples in Modcloth from (a) different models or (b) one model trained with two different random seeds. NMF and LGN is short for NeuMF and LightGCN respectively.}
\vspace{-0.3cm}
\label{fig:prediction_differences}
\end{figure}

To illustrate such signal, we conduct an empirical study on the MovieLens-100k dataset\footnote{
\url{https://grouplens.org/datasets/movielens/100k/}}.
Similar observations also exist on other datasets. 
We train four notable recommendation models (i.e. MF \cite{koren2009matrix}, GMF and NeuMF \cite{he2017neural} and LightGCN \cite{he2020lightgcn}) with the binary user-item implicit feedback with the normal training assumption (i.e., all interacted items are considered as positive examples and negative examples are sampled from missing interactions, as shown in Figure\ref{fig:nomral}).

More precisely, we consider interactions whose ratings are 4 and 5 as clean positive examples while interactions whose ratings are 1 and 2 as noisy positive examples. 
The prediction difference between two models is defined as $|I(\delta(y_{ui}))-I(\delta(y'_{ui}))|$, where $y_{ui}$ and $y'_{ui}$ are the predicted scores from two models regarding user $u$ and item $i$, $\delta$ is the sigmoid function. $I(x)$ is an indicator function and is defined as $I(x)=1$ when $x\geq 0.5$ and $I(x)=0$ otherwise. 
We calculate all prediction differences on clean examples and noisy examples respectively and then report the average.

Figure \ref{fig:Differences_in_models} shows the mean prediction differences between two different models on clean positive examples and noisy positive examples. 
It's obvious that the model prediction differences %\zmc{The `prediction differences' is unclear here.} 
on noisy examples are significantly larger than the differences on clean examples. In other words, \emph{different models tend to make agreement predictions for clean examples compared with noisy examples}.  That is to say, different models tend to fit different parts of the corrupted data but clean examples are the robust agreement component that every model attempts to fit. This observation also conforms with the nature of robust learning. 
Besides, we also find that even one model trained on the same dataset with different random seeds tends to make more consistent prediction agreement on clean examples compared with noisy examples, as shown in Figure \ref{fig:Different_random_seeds}. \wy{One intuitive solution to instantiate this observation is the ensemble methods. However, the data encountered during the inference stage is clean. Then ensemble methods will make consistent while corrupted predictions on the test data, since every model is corrupted. Nevertheless, for our method, we will use the signals from different models to refine one model so that this single target model could potentially yield better predictions.}

%% file: 3.method.tex
\allowdisplaybreaks 
% \vspace{-5pt}
\section{METHODOLOGY} % (fold)
\label{sec:method}
In this section, we propose the learning framework DeCA which uses the observations described in section \ref{sec:motivation} as the denoising signals to learn robust implicit recommenders. Before a detailed description of the methods, some notations and problem formulation are given.
\vspace{-10pt}
\subsection{Notations and Problem Formulation}
\label{sub:notations_and_problem_formulation}
%In this section, we first give some notations in our work. Then we will describe our methods in detail.
% Then we will briefly describe the basic routine in recommender system training and demonstrate the binary cross-entropy loss in a probabilistic way so that it can be unified with our methods later. 
We use $u\in\mathcal{U}$ and $i\in \mathcal{I}$ to denote the user and item with $\mathcal{U}$ and $\mathcal{I}$ being the user and item sets respectively.
$\tilde{\textbf{R}} \in \mathbb{R}^{|\mathcal{U}|\times |\mathcal{I}|}$ is the binary matrix for corrupted data. $\tilde{r}_{ui}$ represents the $(u,i)$-th entry of $\tilde{\textbf{R}}$ and is defined as $\tilde{r}_{ui}=1$ if there is an 
% observed interaction %%%
interaction
between user $u$ and item $i$, otherwise $\tilde{r}_{ui}=0$. Due to the existence of noisy examples, the real preference matrix is different from $\tilde{\textbf{R}}$. We use $\textbf{R}\in \mathbb{R}^{|\mathcal{U}|\times |\mathcal{I}|}$ to denote the real preference matrix and $r_{ui}$ is the real preference of user $u$ over item $i$. What we have is the corrupted binary $\tilde{\textbf{R}}$ while the clean $\textbf{R}$ is not accessible.

Formally, we assume $r_{ui}$ is drawn from a Bernoulli distribution:
\begin{equation}
\label{rui-berno}
  r_{ui} \sim Bernoulli(\eta_{ui})\approx Bernoulli(f_\theta(u,i)),
\end{equation} 
where $\eta_{ui}$ describes the probability of positive preference (i.e., $r_{ui}=1$) and is approximated by $f_\theta(u,i)$. $\theta$ denotes the parameters of $f$. We use $f_\theta$ as our target recommender that generates the final recommendation. \wy{Specifically, the output value after sigmoid of the logits from the model is denoted as $f_\theta(u, i)$, which is also the probability that $\eta_{ui}$ equals to 1.} Generally speaking, $f_\theta$ is expected to have high model expressiveness. Since the denoising signal comes from the agreement across different model predictions, we introduce another auxiliary Bernoulli distribution parameterized by $g_\mu(u,i)$,  In this paper, we use a relatively simple model matrix factorization (MF) \cite{koren2009matrix} as the auxiliary model $g_\mu$. Recent research has demonstrated that MF is still one of the most effective models to capture user preference for recommendation \cite{rendle2020neural}. 

Considering both the noisy positive and noisy negative examples, we assume that given the real preference $r_{ui}$, the corrupted binary $\tilde{r}_{ui}$ is also drawn from Bernoulli distributions as
\begin{equation}\label{h_model}
  \begin{array}{c}
  \tilde{r}_{ui}|r_{ui}=0 \sim Bernoulli(h_\phi(u,i))\\
  \tilde{r}_{ui}|r_{ui}=1 \sim Bernoulli(h'_\psi(u,i)),
  \end{array}
\end{equation}
where $h_\phi(u,i)$ and $h'_\psi(u,i)$, parameterized by $\phi$ and $\psi$ respectively, are two models describing the consistency between the real user preference and data observation. We use two MF models as $h$ and $h'$ without special mention.

The task of this paper is, %\textcolor{red}{Our task/aim is}: 
given the corrupted binary data $\tilde{\textbf{R}}$, to infer the real user preference $\textbf{R}$ and its underlying model $f_\theta$, then to use $f_\theta$ to generate recommendation.

\subsection{Denoising with Cross-Model Agreement} % (fold)
\label{sub:denoising_with_variational_inference}
As discussed in section \ref{sec:motivation}, for clean examples which denote the real user preference, different models tend to make more consistent predictions compared with noisy examples. 
For simplicity, in this subsection we use $P(\mathbf{R})$ to denote the real user preference $Bernoulli(\eta)$. $P_f(\textbf{R})$ and $P_g(\textbf{R})$ are used to represent the approximated $Bernoulli(f_\theta)$ and $Bernoulli(g_\mu)$, correspondingly.

Due to the fact that $P_f(\textbf{R})$ and $P_g(\textbf{R})$ both approximate the real user preference $\textbf{R}$, they should remain a relatively small KL-divergence according to section \ref{sec:motivation}, which is formulated as 
\begin{equation}
\label{eq:kl-original}
D[P_g(\textbf{R})||P_f(\textbf{R})]=E_{\textbf{R}\sim P_g}[\log P_g(\textbf{R})-\log P_f(\textbf{R})].
\end{equation}
However, naively optimizing Eq.(\ref{eq:kl-original}) is meaningless since we do not have the supervision signal of $\mathbf{R}$. As a result, we need to introduce supervision signals from the corrupted data observation $\tilde{\textbf{R}}$. Using the Bayes theorem, $P_f(\textbf{R})$ can be approximated as 
\begin{equation}
\label{eq:bayesian}
P_f(\textbf{R})\approx P(\textbf{R})=\frac{P(\tilde{\textbf{R}}) P(\textbf{R}|\tilde{\textbf{R}})}{P(\tilde{\textbf{R}}|\textbf{R})}.
\end{equation}
Combining Eq.(\ref{eq:bayesian}) and Eq.(\ref{eq:kl-original}), we can obtain
\begin{align}
\label{eq:transformation}
   D&[P_g(\textbf{R}) || P_f(\textbf{R})] = E_{\textbf{R} \sim P_g}[\log P_g(\textbf{R}) - \log P_f(\textbf{R})] \nonumber \\
  &\approx E_{\textbf{R} \sim P_g} [\log P_g(\textbf{R}) - \log \frac{P(\tilde{\textbf{R}}) P(\textbf{R}|\tilde{\textbf{R}})}{P(\tilde{\textbf{R}}|\textbf{R})}] \nonumber \\
  &= E_{\textbf{R} \sim P_g}[\log P_g(\textbf{R}) - \log P(\textbf{R}|\tilde{\textbf{R}}) - \log P(\tilde{\textbf{R}}) + \log P(\tilde{\textbf{R}}|\textbf{R})]  \nonumber \\
  &= D[P_g(\textbf{R})||P(\textbf{R}|\tilde{\textbf{R}})] - \log P(\tilde{\textbf{R}}) + E_{\textbf{R} \sim P_g}[\log P(\tilde{\textbf{R}}|\textbf{R})].
\end{align}
We then rearrange the terms in Eq.(\ref{eq:transformation}) and obtain
\begin{align}
  E_{\textbf{R} \sim P_g}[\log P(\tilde{\textbf{R}}|\textbf{R})] &- D[P_g(\textbf{R}) || P_f(\textbf{R})] \nonumber \\
  &= \log P(\tilde{\textbf{R}}) - D[P_g(\textbf{R})||P(\textbf{R}|\tilde{\textbf{R}})]. \label{lower_bound}
\end{align}
We can see the meaning of maximizing the left side of Eq.(\ref{lower_bound}) is maximizing the likelihood of data observation given real user preference (i.e., $\log P(\tilde{\textbf{R}}|\textbf{R})$) and meanwhile minimizing the KL-divergence between two models which both approximate the real user preference (i.e., $D[P_g(\textbf{R}) || P_f(\textbf{R})]$).
Since the KL-divergence $D[P_g(\textbf{R})||P(\textbf{R}|\tilde{\textbf{R}})]$ is larger than zero, the left side of Eq.(\ref{lower_bound}) can also be seen as the lower bound of $\log P(\tilde{\textbf{R}})$. The bound is satisfied only if $P_g(\textbf{R})$ perfectly recovers $P(\textbf{R}|\tilde{\textbf{R}})$, in other words, $P_g(\textbf{R})$ perfectly approximates the real user preference distribution given the corrupted data.

A naive solution is to directly maximize the left side of Eq.(\ref{lower_bound}) with an end-to-end fashion. However, it would not yield satisfactory performance. The reason is that the left side of Eq.(\ref{lower_bound}) is based on the expectation over $P_g$. The learning process is equivalent to training $g_\mu$ with the corrupted data $\tilde{\textbf{R}}$ and then uses $D[P_g(\textbf{R})||P_f(\textbf{R})]$ to transmit the information from $g_\mu$ to $f_\theta$. However, the information is 
corrupted 
and it will affect the performance of our target model $f_\theta$. To fix the problem, we notice that when the training process is converged, two distributions $P_f(\textbf{R})$ and $P_g(\textbf{R})$ would be close to each other.
We can then modify the left side of Eq.(\ref{lower_bound})
as
\begin{align}
  \label{trick}
  E_{\textbf{R} \sim P_f}[\log P(\tilde{\textbf{R}}|&\textbf{R})] - D[P_g(\textbf{R}) || P_f(\textbf{R})] \nonumber \\
  &\approx \log P(\tilde{\textbf{R}}) - D[P_g(\textbf{R})||P(\textbf{R}|\tilde{\textbf{R}})].
\end{align}

Then optimizing the left side of Eq.(\ref{trick}) is actually training $f_\theta$ with the corrupted $\tilde{\textbf{R}}$ and then transmit information to $g_\mu$. However, since $g_\mu$ is relatively simple than $f_\theta$, $g_\mu$ could only fit the robust data component (i.e., clean examples). 
Thus $g_\mu$ would not affect the learning of $f_\theta$ on clean examples, while in the meantime, pull back $f_\theta$ on noisy samples, or in other words, 
% Then $g_\mu$ will in turn help $f_\theta$ to enhance the learning on clean examples while
downgrade the noisy signal. To this end, the denoising objective function can be formulated as
\begin{align}
  \label{L_simple}
  \mathcal{L}=-E_{\textbf{R} \sim P_f}[\log P(\tilde{\textbf{R}}|\textbf{R})]+D[P_g(\textbf{R}) || P_f(\textbf{R})].
\end{align}
Considering that the gradient of $D[P_g||P_f]$ and $D[P_f||P_g]$ to $\theta$ is different, which could slightly affect the performance, we then formulate the final denoising objective function as
\begin{equation}
  \mathcal{L}_{DeCA} = -E_{\textbf{R}\sim P_f}[\log P(\tilde{\textbf{R}}|\textbf{R})] + \alpha D[P_g||P_f] + (1-\alpha)D[P_f||P_g],
  \label{DPI-loss}
\end{equation}
where $\alpha \in [0,1]$ is a hyper-parameter.

\wy{
In fact, the two models $f_\theta$ and $g_\mu$ in Eq.(\ref{DPI-loss}) can play a symmetric role, but we still need a model (here we use $f_\theta$) to do the inference. Thus, the other model (here we use $g_\mu$) serves as an auxiliary model. For Co-trained DeCA, we chose $g_\mu$ to be MF to (1)  achieve more efficient computation and more stable learning because of the simplicity of MF; (b) yield good performances since MF is shown to be still effective (as stated in Section \ref{sub:notations_and_problem_formulation}); (c) make the proposed DeCA more applicable without tuning on the selection of $g_\mu$, as MF is one of the most general models. 
}
More precisely, we can give the detailed formulation of the terms of \ref{DPI-loss} in Supplement \ref{ssub:mathematical_formulation}.

\subsection{DeCA with Fixed Pre-training}
As described in subsection \ref{sub:denoising_with_variational_inference}, DeCA utilizes an auxiliary model $g_\mu$ as an information filter which helps to downgrade the effect of noisy signal. The auxiliary model $g_\mu$ is co-trained jointly with the target model $f_\theta$. As discussed before, the left side of Eq.(\ref{lower_bound}) is the lower bound of $\log P(\tilde{\textbf{R}})$, which is satisfied when $P_g(\mathbf{R}) \approx P(\textbf{R}|\tilde{\textbf{R}})$. Then as the training process converge, we have $P_g(\mathbf{R}) \approx P_f(\mathbf{R}) $. Thus we could expect $P(\textbf{R}|\tilde{\textbf{R}}) \approx P_f(\mathbf{R})$.

To this end, we modify the assumption of Eq.(\ref{rui-berno}) as 
\begin{equation}
\label{rui|r-berno}
  r_{ui}|\tilde{r}_{ui} \sim Bernoulli(\eta_{ui})\approx Bernoulli(f_\theta(u,i)).
\end{equation} 
The underlying intuition is that whether the data observation is known or not should not affect the robust real user preference, which is reasonable and keeps inline with the nature of robust learning. 
As a result, we should minimize the following KL-divergence
\begin{equation}
    D[P_f(\textbf{R}|\tilde{\textbf{R}}) || P(\textbf{R}|\tilde{\textbf{R}})].
\end{equation}
Through Bayesian transformation, the following equation can be deduced:
\begin{align}
\label{VAE_lower_bound}
E_{P_f}[\log P(\tilde{\textbf{R}}|\textbf{R}&)] - D[P_f(\textbf{R}|\tilde{\textbf{R}})|| P(\textbf{R})] \nonumber \\
  &= \log P(\tilde{\textbf{R}}) - D[P_f(\textbf{R}|\tilde{\textbf{R}})||P(\textbf{R}|\tilde{\textbf{R}})].
\end{align}
Similarly with Eq.(\ref{DPI-loss}), we propose another variant loss function as DeCA with pre-training (DeCA(p)): 
\begin{align}
\label{DVAE-loss}
  \mathcal{L}_{DeCA(p)} &= - E_{P_f}[\log P(\tilde{\textbf{R}}|\textbf{R})] + \alpha D[P_f(\textbf{R}|\tilde{\textbf{R}}) || P(\textbf{R})] \nonumber \\
   &+ (1- \alpha)D[P(\textbf{R}) || P_f(\textbf{R}|\tilde{\textbf{R}})].
\end{align}
We use a pre-trained and fixed model $f_{\theta'}$ which has the same structure as our target model $f_\theta$ but is trained with different random seeds to model $P(\mathbf{R})$. This setting is motivated by the observation that one model trained with different random seeds tends to make high variance predictions on noisy examples but more consistent agreement predictions on clean examples, as shown in Figure \ref{fig:Different_random_seeds}. 

The major difference between DeCA and DeCA(p) can be summarized as follows:
\begin{itemize}[leftmargin=*]
  \item The auxiliary model $g_\mu$ of DeCA is a simple MF while the involved $f_{\theta'}$ in DeCA(p) has the same structure as the target model.
  \item DeCA co-trains the auxiliary model $g_\mu$ with $f_\theta$ but DeCA(p) uses a pre-trained and fixed $f_{\theta'}$ to describe the prior distribution.
\end{itemize}

\subsection{Training Routine} % (fold)
\label{sub:training_routine}
We can see that the objective functions of both DeCA and DeCA(p) contain the term of $E[\log P(\tilde{\textbf{R}}|\textbf{R})]$.
A naive solution to optimize $E[\log P(\tilde{\textbf{R}}|\textbf{R})]$ is direct performing calculation as shown in Eq.(\ref{eq:likelihood}).
However, we find that it won't yield satisfactory results. 
\wy{
The reason is that computing $E[\log P(\tilde{\textbf{R}}|\textbf{R})]$ involves updating three models $h_\phi$, $h'_\psi$ and $f_\theta$ simultaneously. Then these models will interfere with each other. The corrupted outputs from $h_\phi$ and $h'_\psi$ will prevent $f_\theta$ from learning well, and vice versa. 
}
To handle this issue, we split the denoising problem into two dual tasks: \emph{Denoising Positive} (DP) which aims to denoise the noisy positive examples from interacted samples as shown in Figure \ref{fig:FN}; and \emph{Denoising Negative} (DN) which aims to denoise noisy negative examples from sampled missing interactions as illustrated in Figure \ref{fig:FP}. We then optimize these two sub-tasks iteratively. 
\wy{Each time we only consider either $h_\phi$ or $h'_\psi$, and fix the other one to be the function of an approximate large value denoted as $C_1$ and $C_2$ as illustrated in the following part. Then we can obtain information from the data to update $f_\theta$, and also refine one of the $h$-models according to the outputs of $f_\theta$.}
 
\subsubsection{Denoising Positive} % (fold)
\label{ssub:find_negative}
In this situation, we only focus on denoising  noisy positive examples, which means we will regard every negative sample as a clean example,
as shown in Figure \ref{fig:FN}. That is to say,
given $r_{ui} = 1$ we will always have $\tilde{r}_{ui}=1$ (i.e., $h'_\psi(u,i)=1$). %Thus we don't need to care about the distribution $\tilde{r}_{ui}|r_{ui}=1$ in \ref{h_model}, or we freeze $h_\psi'(u,i)$ to be 1. 
Thus in this sub-task, only the model $h_\phi$ will be trained.
Then the term $E[\log P(\tilde{\textbf{R}}|\textbf{R})]$ becomes
\begin{align}
\label{eq:likelihood-dfp}
  &E[\log P(\tilde{\textbf{R}}|\textbf{R})]=\hspace{-0.3cm}\sum_{(u,i)|\tilde{r}_{ui}=1}\hspace{-0.3cm}
     \log h_\phi(u,i) \cdot (1-f_\theta(u,i))\nonumber \\
   &-\hspace{-0.3cm}\sum_{(u,i)|\tilde{r}_{ui}=0}\hspace{-0.3cm}
  C_1\cdot f_{\theta}(u,i) + \log(1 - h_\phi(u,i) ) \cdot (1-f_\theta(u,i)),
\end{align}
where $C_1$ is a large positive hyperparameter to substitute $-\log (1-h'_\psi(u,i))$.
% subsubsection find_negative (end)

\subsubsection{Denoising Negative} % (fold)
\label{ssub:find_positive}
In this sub-task, we only focus on denoising noisy negative examples in the sampled missing interactions, which means that we will regard every positive examples as a clean example, as shown in Figure \ref{fig:FP}.  That is to say, given $r_{ui} = 0$ we will always have $\tilde{r}_{ui}=0$ (i.e., $h_\phi(u,i)=0$). Thus in this sub-task, only model $h'_\psi$ will be trained. The term $E[\log P(\tilde{\textbf{R}}|\textbf{R})]$ is expanded as
\begin{align}
\label{eq:likelihood-dfn}
  &E[\log P(\tilde{\textbf{R}}|\textbf{R})]=\hspace{-0.3cm}\sum_{(u,i)|\tilde{r}_{ui}=0}\hspace{-0.3cm}
   \log(1 - h'_\psi(u,i))\cdot f_{\theta}(u,i) \nonumber \\
 &+\hspace{-0.3cm}\sum_{(u,i)|\tilde{r}_{ui}=1}\hspace{-0.3cm}
    \log h'_\psi(u,i) \cdot f_{\theta}(u,i) -C_2 \cdot (1-f_\theta(u,i)),
 \end{align}
where $C_2$ is a large positive hyperparameter to replace $-\log(h_\phi(u,i))$.
% subsubsection find_positive (end)

In the training procedure of DeCA and DeCA(p), we will expand $E[\log P(\tilde{\textbf{R}}|\textbf{R})]$ as Eq.(\ref{eq:likelihood-dfp}) and Eq.(\ref{eq:likelihood-dfn}) iteratively to denoise both noisy positive examples and noisy negative examples. The detailed training pseudo-code of DeCA and DeCA(p) can be found in the supplement.

% subsection denoising_with_variational_inference (end)

%% file: 4.experiments.tex
\begin{table}
\begin{center}
\caption{Statistics of the datasets}
\vspace{-7pt}
\label{tab:dataset statistics}
\begin{tabular}{c|c|c|c|c}
  \toprule
  Dataset & \# Users & \# Items & \# Interactions & Sparsity \\
  \midrule
  MovieLens & 943 & 1,682 & 100,000 & 0.93695 \\
%   \hline
  Modcloth & 44,783 & 1,020 & 99,893 & 0.99781 \\
%   \hline 
  Adressa & 212,231 & 6,596 & 419,491 & 0.99970 \\
%   \hline
  Electronics & 1,157,633 & 9,560 & 1,292,954 & 0.99988 \\
  \bottomrule
\end{tabular}
\end{center}
\vspace{-0.5cm}
\end{table}

\section{EXPERIMENTS} % (fold)
\label{sec:experiments}

\begin{table*}[ht]
\centering
\footnotesize
\caption{Overall performance comparison. The highest scores are in Boldface. R is short for Recall and N is short for NDCG. \wy{The results with improvements over the best baseline larger than 5\% are marked with $*$}.}
\vspace{-0.3cm}
\label{tab:Overall_performance}
\resizebox{\textwidth}{!}{%
\begin{tabular}{c|cccc|cccc}
\hline
\multirow{2}{*}{\textbf{MovieLens}}& R@3 & R@20 &N@3 &N@20 & R@3 & R@20 &N@3 &N@20 \\
\cmidrule(lr){2-5}\cmidrule{6-9}
 &\multicolumn{4}{c|}{GMF} & \multicolumn{4}{c}{NeuMF} \\ 
    \hline
Normal & 0.021$\pm$0.001 & 0.095$\pm$0.001 & 0.035$\pm$0.003 & 0.059$\pm$0.001 & 0.025$\pm$0.002 & 0.103$\pm$0.006 & 0.041$\pm$0.002 & 0.064$\pm$0.003 \\  
WBPR & 0.023$\pm$0.002 & 0.082$\pm$0.003 & 0.045$\pm$0.002 & 0.056$\pm$0.000 & 0.022$\pm$0.003 & 0.086$\pm$0.001 & 0.038$\pm$0.005 & 0.054$\pm$0.002 \\ 
T-CE & 0.017$\pm$0.002 & 0.098$\pm$0.001 & 0.026$\pm$0.003 & 0.054$\pm$0.001 &  0.026$\pm$0.004 & 0.106$\pm$0.002 & 0.047$\pm$0.003 & 0.068$\pm$0.002 \\ 
Ensemble & 0.021$\pm$0.000 & 0.095$\pm$0.001 & 0.035$\pm$0.001 & 0.059$\pm$0.000 & 0.030$\pm$0.001 & 0.108$\pm$0.003 & 0.048$\pm$0.002 & 0.070$\pm$0.002 \\ 
DeCA & 0.022$\pm$0.001 & \textbf{0.109$\pm$0.007$^*$}
& 0.038$\pm$0.001 & 0.064$\pm$0.002 & 0.022$\pm$0.001 & 0.099$\pm$0.006 & 0.032$\pm$0.003 & 0.059$\pm$0.003 \\ 
DeCA(p) & \textbf{0.027$\pm$0.002$^*$} & 0.099$\pm$0.001 & \textbf{0.054$\pm$0.003$^*$} & \textbf{0.066$\pm$0.001$^*$} & \textbf{0.035$\pm$0.000$^*$} & \textbf{0.120$\pm$0.003$^*$} & \textbf{0.064$\pm$0.000$^*$} & \textbf{0.081$\pm$0.002$^*$} \\ 
\hline
& \multicolumn{4}{c|}{CDAE} & \multicolumn{4}{c}{LightGCN} \\ 
\hline
Normal & 0.017$\pm$0.000 & 0.095$\pm$0.002 & 0.028$\pm$0.004 & 0.052$\pm$0.001 & 0.025$\pm$0.000 & 0.106$\pm$0.001 & 0.048$\pm$0.000 & 0.065$\pm$0.000 \\
WBPR & 0.017$\pm$0.003 & 0.087$\pm$0.004 & 0.028$\pm$0.003 & 0.050$\pm$0.002 & \textbf{0.031}$\pm$0.000 & 0.093$\pm$0.001 & 0.054$\pm$0.000 & 0.064$\pm$0.000 \\ 
T-CE & 0.014$\pm$0.002 & 0.095$\pm$0.007 & 0.024$\pm$0.004 & 0.050$\pm$0.001 & 0.011$\pm$0.001 & 0.080$\pm$0.007 & 0.020$\pm$0.002 & 0.043$\pm$0.002 \\  
Ensemble & 0.015$\pm$0.003 & 0.099$\pm$0.000 & 0.026$\pm$0.004 & 0.052$\pm$0.001 & 0.026$\pm$0.001 & 0.105$\pm$0.002 & 0.049$\pm$0.001 & 0.064$\pm$0.000 \\ 
DeCA & \textbf{0.029$\pm$0.001$^*$} & 0.109$\pm$0.000 & \textbf{0.048$\pm$0.001$^*$} & 0.066$\pm$0.000 & 0.027$\pm$0.000 & 0.104$\pm$0.001 & 0.044$\pm$0.002 & 0.064$\pm$0.001 \\  
DeCA(p) & 0.028$\pm$0.002 & \textbf{0.110$\pm$0.002$^*$} & 0.047$\pm$0.002 & \textbf{0.067$\pm$0.001$^*$} & 0.029$\pm$0.001 & \textbf{0.118$\pm$0.001$^*$} & \textbf{0.055}$\pm$0.001 & \textbf{0.075$\pm$0.001$^*$} \\ 
\hline
\hline
\multirow{2}{*}{\textbf{Modcloth}}& R@5 & R@20 &N@5 &N@20 & R@5 & R@20 &N@5 &N@20 \\
\cmidrule(lr){2-5}\cmidrule{6-9}
 &\multicolumn{4}{c|}{GMF} & \multicolumn{4}{c}{NeuMF} \\ 
    \hline
Normal & 0.063$\pm$0.006 & 0.225$\pm$0.008 & 0.043$\pm$0.005 & 0.088$\pm$0.005 & 0.082$\pm$0.003 & 0.242$\pm$0.004 & 0.055$\pm$0.002 & 0.101$\pm$0.002 \\
WBPR & 0.067$\pm$0.002 & 0.224$\pm$0.002 & 0.046$\pm$0.001 & 0.090$\pm$0.001 & 0.092$\pm$0.001 & 0.247$\pm$0.020 & 0.064$\pm$0.001 & 0.108$\pm$0.006 \\
T-CE & 0.067$\pm$0.004 & 0.235$\pm$0.002 & 0.045$\pm$0.003 & 0.093$\pm$0.001 & 0.065$\pm$0.010 & 0.228$\pm$0.008 & 0.044$\pm$0.010 & 0.091$\pm$0.008 \\
Ensemble & 0.064$\pm$0.004 & 0.228$\pm$0.007 & 0.043$\pm$0.002 & 0.089$\pm$0.002 & 0.090$\pm$0.003 & 0.260$\pm$0.004 & 0.061$\pm$0.001 & 0.109$\pm$0.002 \\ 
DeCA & 0.072$\pm$0.001 & 0.245$\pm$0.001 & 0.050$\pm$0.001 & 0.099$\pm$0.001 & \textbf{0.099$\pm$0.005$^*$} & \textbf{0.268}$\pm$0.006 & \textbf{0.065}$\pm$0.001 & \textbf{0.113}$\pm$0.001 \\
DeCA(p) & \textbf{0.074$\pm$0.001$^*$} & \textbf{0.247$\pm$0.001$^*$} & \textbf{0.052$\pm$0.001$^*$} & \textbf{0.100$\pm$0.000$^*$} & 0.087$\pm$0.005 & 0.265$\pm$0.006 & 0.059$\pm$0.004 & 0.110$\pm$0.004 \\
\hline
& \multicolumn{4}{c|}{CDAE} & \multicolumn{4}{c}{LightGCN} \\ 
\hline
Normal & 0.082$\pm$0.004 & 0.242$\pm$0.003 & 0.052$\pm$0.002 & 0.098$\pm$0.001 & 0.065$\pm$0.001 & 0.220$\pm$0.002 & 0.043$\pm$0.001 & 0.087$\pm$0.002 \\
WBPR & 0.079$\pm$0.000 & 0.238$\pm$0.004 & 0.050$\pm$0.000 & 0.095$\pm$0.001 & 0.072$\pm$0.001 & 0.222$\pm$0.002 & 0.046$\pm$0.001 & 0.088$\pm$0.001 \\
T-CE & 0.075$\pm$0.003 & 0.243$\pm$0.008 & 0.048$\pm$0.002 & 0.096$\pm$0.002 & 0.071$\pm$0.000 & 0.231$\pm$0.001 & 0.049$\pm$0.000 & 0.093$\pm$0.000 \\
Ensemble & 0.084$\pm$0.002 & 0.250$\pm$0.002 & 0.054$\pm$0.001 & 0.100$\pm$0.001 & 0.068$\pm$0.001 & 0.227$\pm$0.001 & 0.046$\pm$0.000 & 0.090$\pm$0.000 \\ 
DeCA & 0.086$\pm$0.004 & 0.250$\pm$0.003 & 0.056$\pm$0.001 & 0.102$\pm$0.000 & 0.064$\pm$0.001 & 0.221$\pm$0.002 & 0.041$\pm$0.000 & 0.085$\pm$0.000 \\
DeCA(p) & \textbf{0.089$\pm$0.004$^*$} & \textbf{0.251}$\pm$0.005 & \textbf{0.057$\pm$0.003$^*$} & \textbf{0.103$\pm$0.003$^*$} & \textbf{0.073$\pm$0.001} & \textbf{0.235$\pm$0.001} & \textbf{0.051$\pm$0.001} & \textbf{0.096$\pm$0.000} \\
\hline
\hline
\multirow{2}{*}{
    \textbf{Adressa}} & R@5 & R@20 &N@5 &N@20 & R@5 & R@20 &N@5 &N@20 \\
\cmidrule(lr){2-5}\cmidrule{6-9} & \multicolumn{4}{c|}{GMF} & \multicolumn{4}{c}{NeuMF} \\
    \hline
Normal&0.116$\pm$0.003&0.209$\pm$0.005&0.080$\pm$0.002&0.112$\pm$0.001&0.169$\pm$0.004&0.312$\pm$0.004&0.131$\pm$0.002&0.180$\pm$0.003 \\
WBPR&0.115$\pm$0.007&0.210$\pm$0.007&0.084$\pm$0.003&0.116$\pm$0.003&0.172$\pm$0.002&0.311$\pm$0.003&0.132$\pm$0.001&0.181$\pm$0.001 \\
T-CE&0.109$\pm$0.000&0.209$\pm$0.001&0.070$\pm$0.000&0.104$\pm$0.000&0.172$\pm$0.003&0.312$\pm$0.003&0.134$\pm$0.001&\textbf{0.183$\pm$0.002} \\
Ensemble & 0.110$\pm$0.002 & 0.191$\pm$0.006 & 0.078$\pm$0.002 & 0.105$\pm$0.003 & 0.180$\pm$0.001 & 0.311$\pm$0.001 & 0.134$\pm$0.001 & 0.180$\pm$0.000   \\
DeCA&\textbf{0.125$\pm$0.002$^*$} &\textbf{0.220$\pm$0.001$^*$} & 0.091$\pm$0.002 & 0.126$\pm$0.002 & 0.170$\pm$0.006 & \textbf{0.318$\pm$0.002} & 0.130$\pm$0.001 & 0.181$\pm$0.000 \\
DeCA(p)& 0.123$\pm$0.005 &  0.220$\pm$0.001 & \textbf{0.093$\pm$0.004$^*$} & \textbf{0.127$\pm$0.003$^*$} & \textbf{0.183$\pm$0.009}&0.316$\pm$0.004&\textbf{0.137$\pm$0.002}&0.181$\pm$0.005 \\
\hline
  & \multicolumn{4}{c|}{CDAE} & \multicolumn{4}{c}{LightGCN} \\ 
\hline
Normal & 0.162$\pm$0.000 & 0.317$\pm$0.001 & 0.123$\pm$0.000 & 0.178$\pm$0.000 & 0.085$\pm$0.004 & 0.215$\pm$0.005 & 0.064$\pm$0.003 & 0.107$\pm$0.003 \\
WBPR & 0.161$\pm$0.001 & 0.315$\pm$0.005 & 0.121$\pm$0.002 & 0.173$\pm$0.004 & 0.118$\pm$0.003 & 0.211$\pm$0.006 & 0.089$\pm$0.002 & 0.119$\pm$0.004 \\
T-CE & 0.161$\pm$0.001 & 0.317$\pm$0.003 & 0.122$\pm$0.002 & 0.176$\pm$0.004 & 0.119$\pm$0.001 & 0.206$\pm$0.003 & \textbf{0.091}$\pm$0.001 & 0.121$\pm$0.001 \\
Ensemble &  0.162$\pm$0.000 & 0.317$\pm$0.001 & 0.122$\pm$0.000 & 0.176$\pm$0.000 & 0.104$\pm$0.001 & 0.217$\pm$0.001 & 0.078$\pm$0.001 & 0.118$\pm$0.001 \\
DeCA & 0.162$\pm$0.000 & 0.319$\pm$0.000 & 0.123$\pm$0.000 & 0.178$\pm$0.000 & 0.112$\pm$0.001 & 0.221$\pm$0.001 & 0.077$\pm$0.005 & 0.116$\pm$0.005 \\
DeCA(p) & \textbf{0.163$\pm$0.000} & \textbf{0.320$\pm$0.002} & \textbf{0.123$\pm$0.000} & \textbf{0.178$\pm$0.001} & \textbf{0.121$\pm$0.001} & \textbf{0.222$\pm$0.001} & 0.089$\pm$0.001 & \textbf{0.125$\pm$0.000} \\
\hline
\end{tabular}}
\vspace{-0.2cm}
% \vspace{-5pt}
\end{table*}

In this section, we instantiate the proposed DeCA and DeCA(p) with four state-of-the-art recommendation models as the target model $f$ and conduct experiments on four real-world datasets to verify the effectiveness of our methods. We aim to answer the following research questions:
% \href{}{GoogleDrive}
\begin{itemize}[leftmargin=*]
  \item \textbf{RQ1}: How do the proposed methods perform compared to normal training and other denoising methods? Can the proposed methods help to downgrade the effect of noisy examples?
  \item \textbf{RQ2}: How does the design of proposed methods affect the recommendation performance, including the iterative training, hyperparameter study, and model selection?
  \item \textbf{RQ3}: Can the proposed methods generate reasonable preference distribution given the corrupted binary data observation?
\end{itemize}
\subsection{Experimental Settings} % (fold)
\label{sub:experimental_settings}
\subsubsection{Datasets} % (fold)
\label{ssub:datasets}
We conduct experiments with four public accessible datasets: MovieLens, Modcloth\footnote{https://github.com/MengtingWan/marketBias}, Adressa\footnote{https://github.com/WenjieWWJ/DenoisingRec} and Electronics\footnote{https://github.com/MengtingWan/marketBias}. 
More statistics about these three datasets are listed in Table \ref{tab:dataset statistics}. For each dataset, we construct the clean test set with only clean examples which denote the real user preference.

\textbf{MovieLens}~\cite{DBLP:journals/tiis/HarperK16}. This is one of the most popular datasets in the task of recommendation. For evaluation, the clean test set is constructed based on user-item pairs with ratings equal to 5. 

\textbf{Modcloth}. It is from an e-commerce website that sells women's clothing and accessories. For evaluation, the clean test set is built on user-item pairs whose rating scores are equal to 5.

\textbf{Adressa}. This is a real-world news reading dataset from Adressavisen. It contains the interaction records of anonymous users and the news. %The dwell time for each user-item pair is used to construct the clean test set. 
According to \cite{kim2014modeling}, we use interactions with dwell time longer than 10 seconds to construct the clean test set.

\textbf{Electronics}. Electronics is collected from the \emph{Electronics} category on Amazon \cite{ni2019justifying,wan2020addressing}. %It is based on the public \emph{Amazon 2018 Dataset}\cite{ni2019justifying} and further processed by \cite{wan2020addressing}. 
The clean test set is built with user-item pairs with rating scores equal to 5.

Note that all the ratings and dwell time are only used to construct the clean test set. The models are trained with only the corrupted binary implicit feedback.

\subsubsection{Evaluation protocols} % (fold)
\label{ssub:evaluation_protocals}
We adopt cross-validation to evaluate the performance. 
For Adressa and MovieLens, we split the user-item interactions into the training set, validation set, and test set according to the ratio of 8:1:1 in chronological order \cite{wang2020denoising}. As for the other two datasets, we randomly split the historical interactions according to the ratio of 8:1:1. After that, the clean test set is constructed for each dataset as described in section \ref{ssub:datasets}.
The performance is measured by two widely used top-$K$ recommendation metrics \cite{he2017neural, yang2018hop}: recall@$K$ and ndcg@$K$. By default, we set K = 5, 20 for Modcloth and Adressa, and K = 10, 50 for Electronics since the number of items in Electronics is larger. All experiments are run 3 times. The average and standard deviation are reported.
\subsubsection{Baselines} % (fold)
\label{ssub:baselines}
We select four state-of-the-art recommendation models as the target model $f$ of DeCA and DeCA(p):
\begin{itemize}[leftmargin=*]
  \item \textbf{GMF}~\cite{he2017neural}: This is a generalized version of MF by changing the inner product to the element-wise product and a dense layer.
  \item \textbf{NeuMF}~\cite{he2017neural}: The method is a state-of-the-art neural CF model which combines GMF with a Multi-Layer Perceptron (MLP).
  \item \textbf{CDAE}~\cite{wu2016collaborative}: CDAE corrupts the 
  observed interactions %%% 
  with random noises, and then employs several linear layers to reconstruct the original datasets, which will increase its anti-noise abilities.
  \item \textbf{LightGCN}~\cite{he2020lightgcn}: LightGCN is a newly proposed graph-based recommendation model which learns user and item embeddings by linearly propagating them on the interaction graph.% and uses the weighted sum of the embeddings learned at all layers as the final embedding, which is easy to implement and train, yet with excellent performance.
\end{itemize}
Each model is trained with the following approaches:
\begin{itemize}[leftmargin=*]
  \item \textbf{Normal}: Train the model with the assumption shown in Figure \ref{fig:nomral} and simple binary-cross-entropy (BCE) loss.
  \item \textbf{WBPR}~\cite{gantner2012personalized}: This is a re-sampling based denoising method, which considers the popular but uninteracted items are highly likely to be real negative ones.
  \item \textbf{T-CE}~\cite{wang2020denoising}: This is a re-weighting based denoising method, which uses the Truncated BCE to assign zero weights to large-loss examples with a dynamic threshold in each iteration.
  \item \textbf{Ensemble}: This is an ensemble-based approach, which aggregates the results from two models with different random seeds.
%   examples with high loss values 
  \item \textbf{DeCA} and \textbf{DeCA(p)}: Our proposed methods.
\end{itemize}
Besides, \cite{yu2020sampler} proposed a noisy robust learning method. We do not compare with this method because it has been shown to be only applicable to the MF model \cite{yu2020sampler}.
% subsubsection baselines (end)

\subsubsection{Parameter settings} % (fold)
\label{ssub:parameter_settings}
We optimize all models using Adam optimizer. The batch size is set as 2048. The learning rate is tuned as 0.001 on four datasets.
For each training instance, we sample one interacted sample and one randomly sampled missing interaction to feed the model. 
We use the recommended network settings of all models.
Besides, the embedding size of users and items in GMF and NeuMF, the hidden size of CDAE, are all set to 32. 
For LightGCN, the embedding size is set to 64 without dropout \cite{he2020lightgcn}. The $L_2$ regularization coefficient is tuned in $\{0.01, 0.1, 1, 10, 100, 1000\}$ divided by the number of users in each dataset.
For DeCA and DeCA(p), there are three hyper-parameters: $C_1$, $C_2$ and $\alpha$. 
We apply a grid search for hyperparameters: $C_1$ and $C_2$ are searched among \{1, 10, 100, 1000\},  $\alpha$ is tuned in \{0, 0.5, 1\}. 
Note that the hyperparameters of recommendation models keep exactly the same across all training approaches for a fair comparison. It also indicates that the proposed methods can be easily integrated with downstream recommendation models without exhaustive hyperparameter refinement.

% subsubsection parameter_settings (end)
% subsection experimental_settings (end)
\vspace{-5pt}
\subsection{Performance Comparison (RQ1)} % (fold)
\label{sub:performance_comparison}
Tabel \ref{tab:Overall_performance} shows the performance comparison on Movielens, Modcloth and Adressa. The results on  Electronics are provided in the supplement due to space limitations. 
From the table, we can have the following observations:
\begin{itemize}[leftmargin=*]
  \item The proposed DeCA and DeCA(p) can effectively improve the recommendation performance of all the four recommendation models over all datasets. Either DeCA or DeCA(p) achieves the best performance compared with normal training and other denoising methods, except for few cases. Even if the CDAE model itself is based on the denoising auto-encoder which is more robust to corrupted data, there still exists improvement when training with our proposed methods, especially in the MovieLens dataset.
  \item Simple ensemble of the results from different models sometimes cannot lead to better performance, for example, in the Electronic dataset. The reason could be that the Electronic data is super sparse. The results from different models vary much more. Simple aggregating two very different results could lead to poor performance in such cases.  
  \item The performance of LightGCN is not so good on Modcloth, Adressa and Electronics. This could be attributed to the bias of datasets.
  We can see from Table \ref{tab:dataset statistics} that users in three datasets would only be connected to a small number of items while items are connected to a large amount of users. The imbalance of interaction graphs could affect the performance of LightGCN.
\end{itemize}

\begin{figure}
\centering     %%% not \center
\subfigure[Modcloth]{\label{fig:GMF_modcloth_recall}\includegraphics[width=0.32\linewidth]{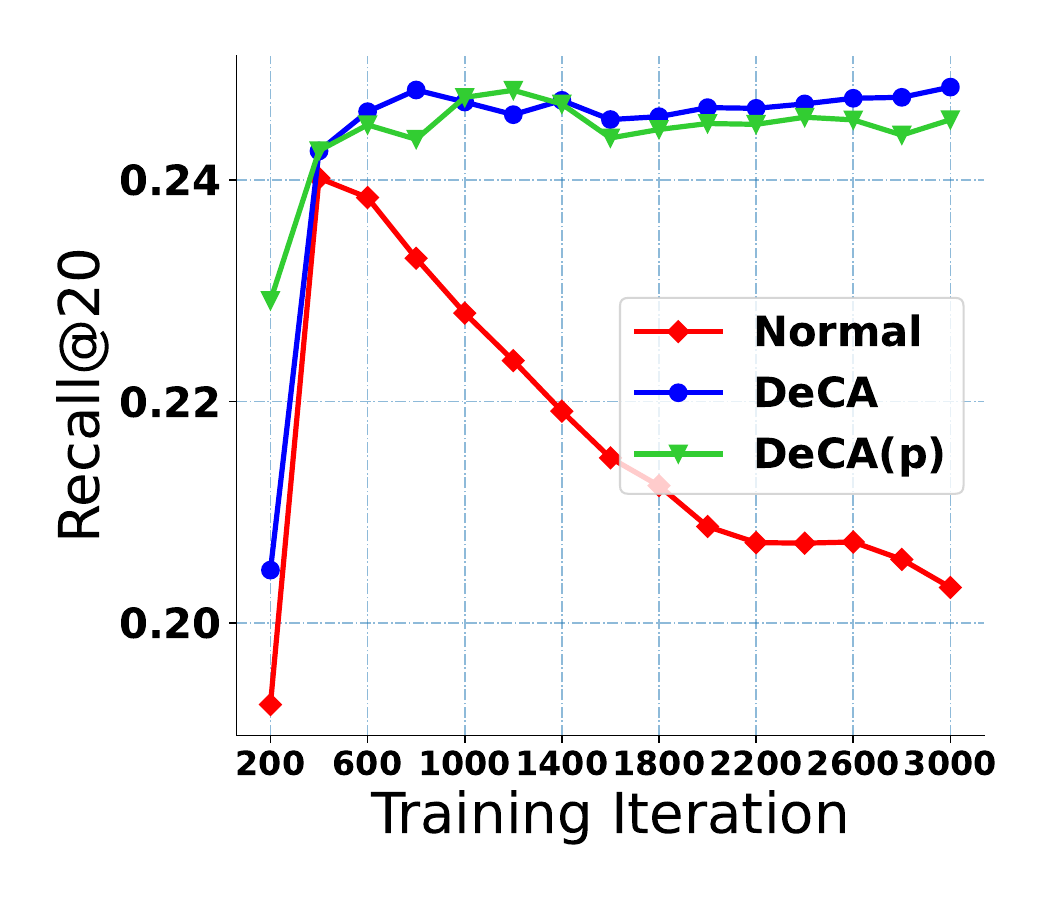}}
\subfigure[Adressa]{\label{fig:GMF_adressa_recall}\includegraphics[width=0.32\linewidth]{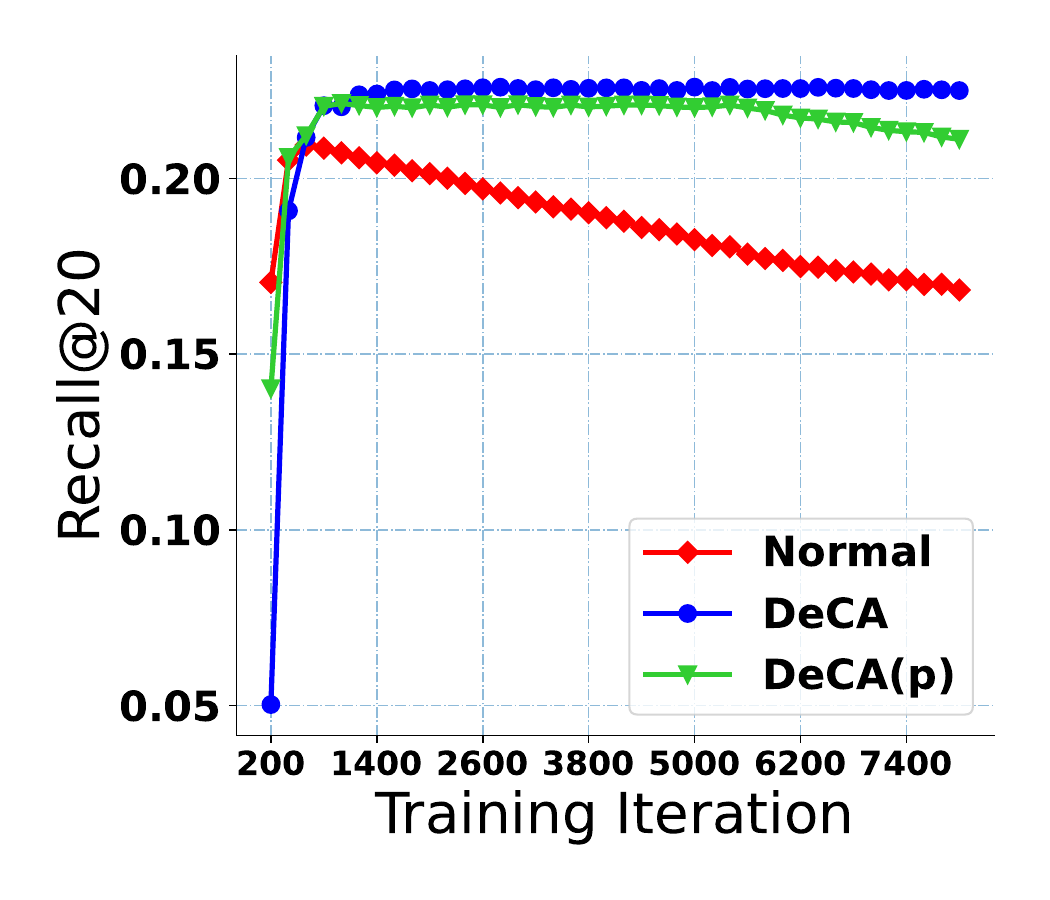}}
\subfigure[Electronics]{\label{fig:GMF_electronics_recall}\includegraphics[width=0.32\linewidth]{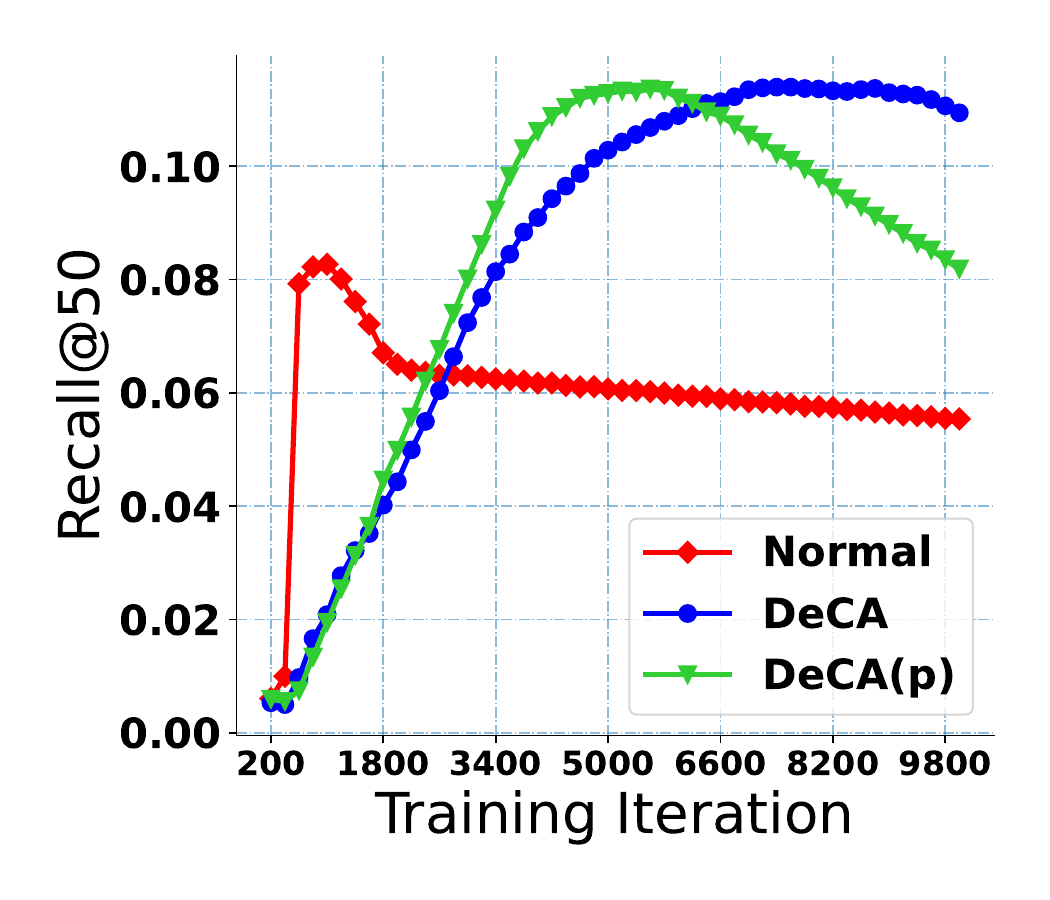}}
\vspace{-0.4cm}
\caption{Recall along the training process}
\vspace{-15pt}
\label{fig:Analysis_on_memorizing_noisy_samples}
\end{figure}

Besides, we also analyse how DeCA and DeCA(p) affect the memorization of noisy samples. Figure \ref{fig:Analysis_on_memorizing_noisy_samples} shows the learning curve of normal training and the proposed DeCA and DeCA(p) on Modcloth, Adressa and Electronics when using GMF as the target recommender. We can see that in normal training, performance on the clean test set decreases in the late training stage. The reason is as the training goes on, the model tends to memorize all samples in the dataset, both noisy and clean ones. However, we can see that both DeCA and DeCA(p) remain more stable and better performance along the whole training stage. The results demonstrate that the proposed DeCA and DeCA(p) successfully prevent the model from being affected by noisy samples. 

\subsection{Model Investigation (RQ2)}
\subsubsection{Ablation Study} % (fold)
\label{ssub:ablation_study}
DeCA and DeCA(p) utilize an iterative training routine that contains two sub-tasks DP and DN. In this part, we discuss how these two sub-tasks contribute to the whole framework and how they perform separately. 
The results on Modcloth dataset are shown in Table \ref{tab:Ablation_study_with_all_models_on_modcloth}. 
We can have following observations: 
\begin{itemize}[leftmargin=*]
\item In most cases, DeCA and DeCA(p) are better than only considering one sub-task of denoising. Some abnormal cases could be attributed to the instability of DN.
During each training epoch, the interacted samples are fixed while the sampled negative instances could change frequently.
% according to the negative sampling strategy
Since DN aims to denoise noisy negative examples from the sampled missing interactions, the performance of DN could be unstable. 
\item DP is better than DN in most cases, except for the LightGCN model. This observation to some extent indicates that finding noisy examples from interacted samples is much easier than finding potential positive preference from the massive uninteracted missing samples.
\end{itemize}

\subsubsection{Hyperparameter Study.} % (fold)
\begin{figure}
\centering     %%% not \center
\vspace{-0.2cm}
\subfigure[DeCA]{\label{fig:hyper_GMF_adressa_DeCA}\includegraphics[width=0.430\linewidth]{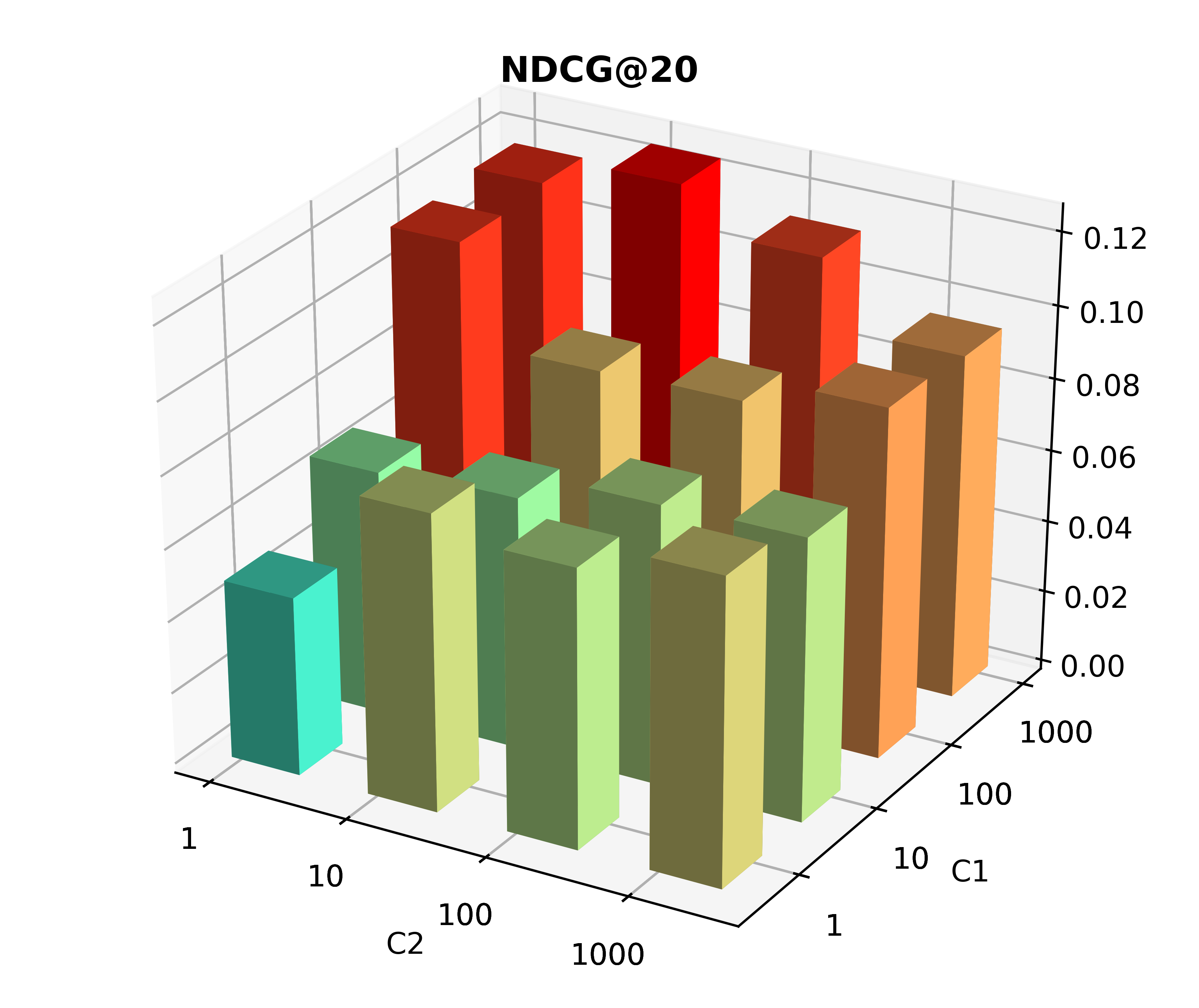}}
% \hfill
\subfigure[DeCA(p)]{\label{fig:hyper_GMF_adressa_DeCA(p)}\includegraphics[width=0.430\linewidth]{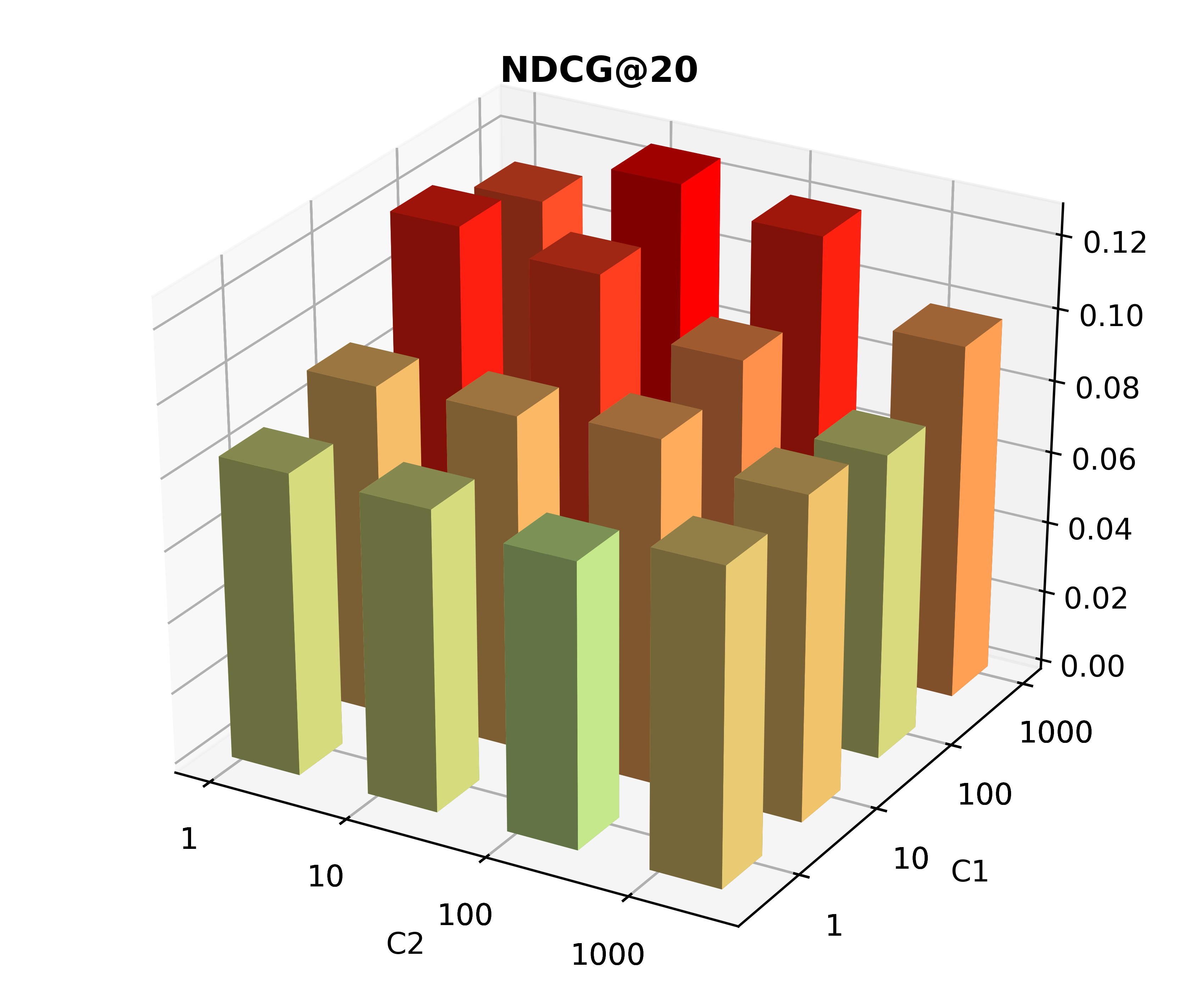}}
\vspace{-0.3cm}
\caption{Hyperparameter study on Adressa.}
\vspace{-0.3cm}
\label{fig:Hyper_parameter_sensitivity}
\end{figure}

\label{ssub:hyper_parameter_sensitivity}
In this part, we conduct experiments to show the effect of $C_1$ and $C_2$, which are two hyper-parameters used in the iterative training routine.
More specifically, a larger $C_1$ means we are more confident that the negative samples are truly negative. A larger $C_2$ denotes that an interacted sample is more likely to be real positive. Figure \ref{fig:Hyper_parameter_sensitivity} shows the results on Adressa when using GMF as the target recommendation model. Results on other datasets are provided in the supplement. 
We find that the optimal setting of $C_1$ (i.e., $C_1=1000$) is larger than the optimal setting of $C_2$ (i.e., $C_2=10 \text{ or } 100$). 
This observation indicates that the probability that a missing interaction is real negative is larger than the probability that an interacted sample is real positive. 

\begin{table*}
\centering
\caption{Performance comparison when considering one sub-task.
DeCA-DP and DeCA-DN denote training DeCA with only either DP or DN. DeCA(p)-DP and DeCA(p)-DN denote training DeCA(p) with only either DP or DN.}
\vspace{-0.3cm}
\label{tab:Ablation_study_with_all_models_on_modcloth}
\resizebox{\textwidth}{!}{%
\begin{tabular}{c|cccc|cccc|cccc|cccc}
  \hline
& \multicolumn{4}{c|}{GMF} & \multicolumn{4}{c|}{NeuMF} & \multicolumn{4}{c|}{CDAE} & \multicolumn{4}{c}{LightGCN} \\ 
& R@5 & R@20 &N@5 &N@20 & R@5 & R@20 &N@5 &N@20 & R@5 & R@20 &N@5 &N@20 & R@5 & R@20 &N@5 &N@20 \\
\hline
\hline
DeCA-DP & 0.072 & 0.245 & 0.050 & 0.098 & 0.089 & 0.256 & 0.061 & 0.109 & 0.077 & 0.236 & 0.050 & 0.095 & 0.066 & 0.223 & 0.042 & 0.086 \\
DeCA-DN & 0.051 & 0.187 & 0.032 & 0.071 & 0.045 & 0.198 & 0.027 & 0.070 & 0.066 & 0.236 & 0.044 & 0.093 & \textbf{0.068} & \textbf{0.240} & \textbf{0.047} & \textbf{0.096} \\
DeCA & \textbf{0.072} & \textbf{0.245} & \textbf{0.050} & \textbf{0.099} & \textbf{0.099} & \textbf{0.268} & \textbf{0.065} & \textbf{0.113} & \textbf{0.086} & \textbf{0.250} & \textbf{0.056} & \textbf{0.102} & 0.064 & 0.221 & 0.041 & 0.085 \\
\hline
DeCA(p)-DP & \textbf{0.075} & 0.243 & 0.052 & 0.099 & \textbf{0.094} & \textbf{0.268} & \textbf{0.064} & \textbf{0.113} & \textbf{0.091} & 0.245 & \textbf{0.059} & 0.102 & 0.065 & 0.220 & 0.042 & 0.086 \\
DeCA(p)-DN & 0.060 & 0.221 & 0.039 & 0.084 & 0.092 & 0.255 & 0.062 & 0.108 & 0.073 & 0.238 & 0.050 & 0.097 & 0.072 & \textbf{0.240} & 0.049 & 0.096 \\
DeCA(p) & 0.074 & \textbf{0.247} & \textbf{0.052} & \textbf{0.100} & 0.087 & 0.265 & 0.059 & 0.110 & 0.089 & \textbf{0.251} & 0.057 & \textbf{0.103} & \textbf{0.073} & 0.235 & \textbf{0.051} & \textbf{0.096} \\
\hline
\end{tabular}}
\end{table*}
% \vspace{-10pt}
% subsubsection hyper_parameter_sensitivity (end)

\subsubsection{Effect of Model Selection} % (fold)
\label{ssub:Effect_of_Model_Selection}
We use MF as $h$ and $h'$ to model the probability $P(\tilde{\textbf{R}}|\textbf{R})$ as our default settings. Since the model capability of MF might be limited, in this part we conduct experiments to see how the performance would be if we use more complicated models for $h$ and $h'$. Table \ref{tab:GMF_modcloth_gamma} shows the result on Modcloth. Results on other datasets are provided in the supplement. The target recommendation model is GMF.  
We can see that replacing MF with more complicated models will not boost the performance significantly. The reason might be that modelling the probability $P(\tilde{\textbf{R}} | \textbf{R})$ is not a very complex task. Accomplishing this task with MF already works well. 
Besides, as we can see that no matter what model we use, the performance of the proposed DeCA and DeCA(p) are significantly better than normal training.

\begin{figure*}[h!]
\subfigure[MovieLens-DeCA]{\label{fig:DPI_movielens_rating}\includegraphics[width=0.16\textwidth]{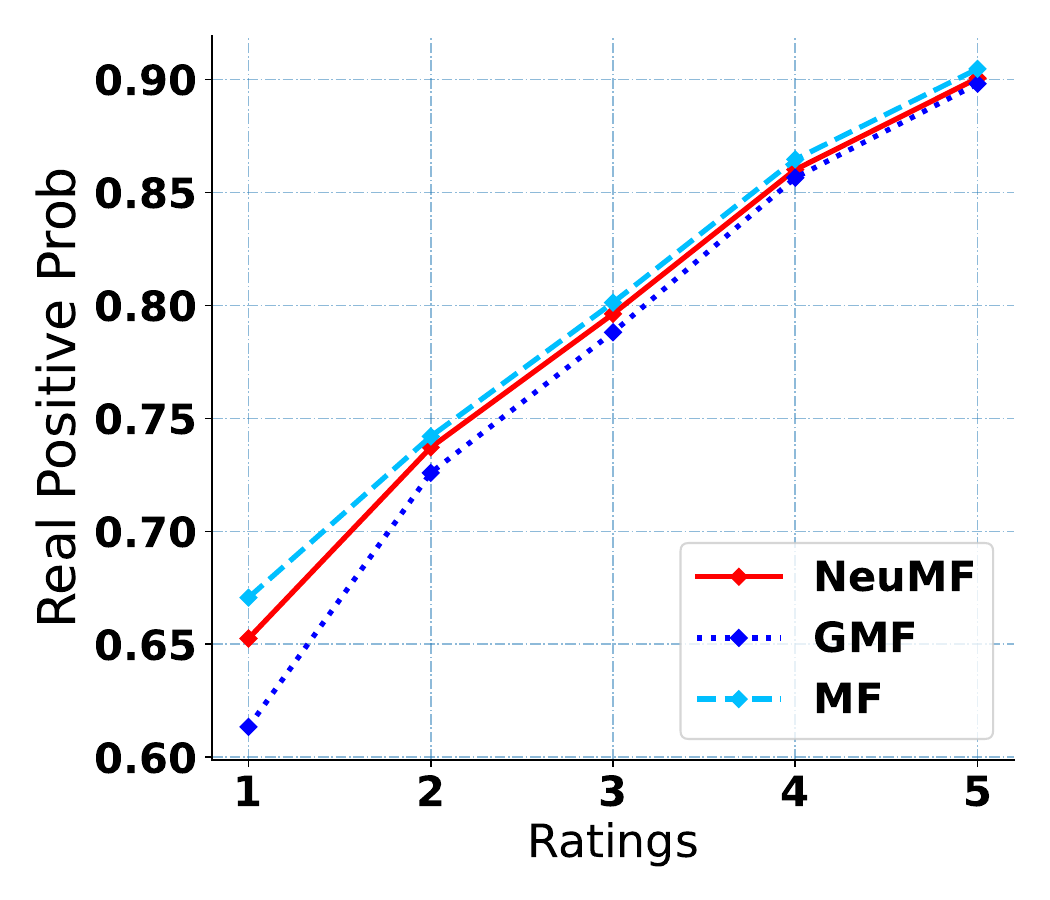}}
\subfigure[MovieLens-DeCA(p)]{\label{fig:DVAE_movielens_rating}\includegraphics[width=0.16\textwidth]{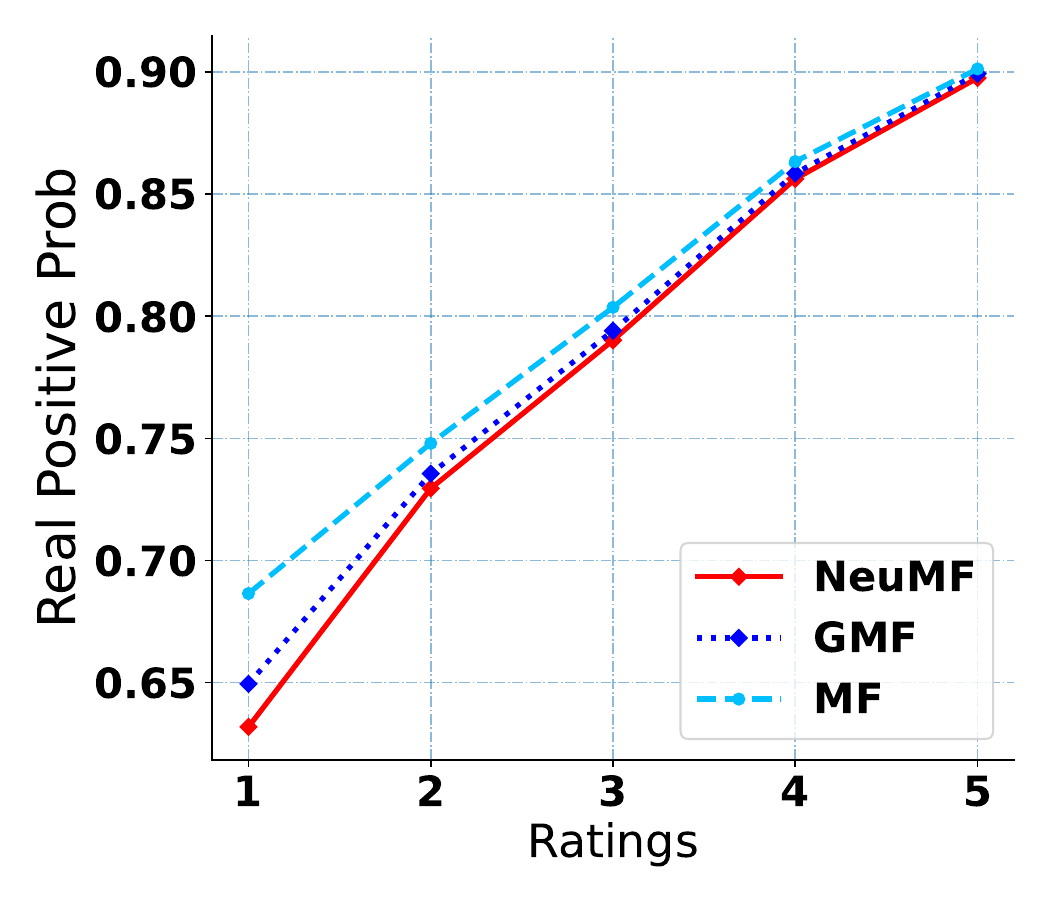}}
\subfigure[Modcloth-DeCA]{\label{fig:DPI_modcloth_rating}\includegraphics[width=0.16\textwidth]{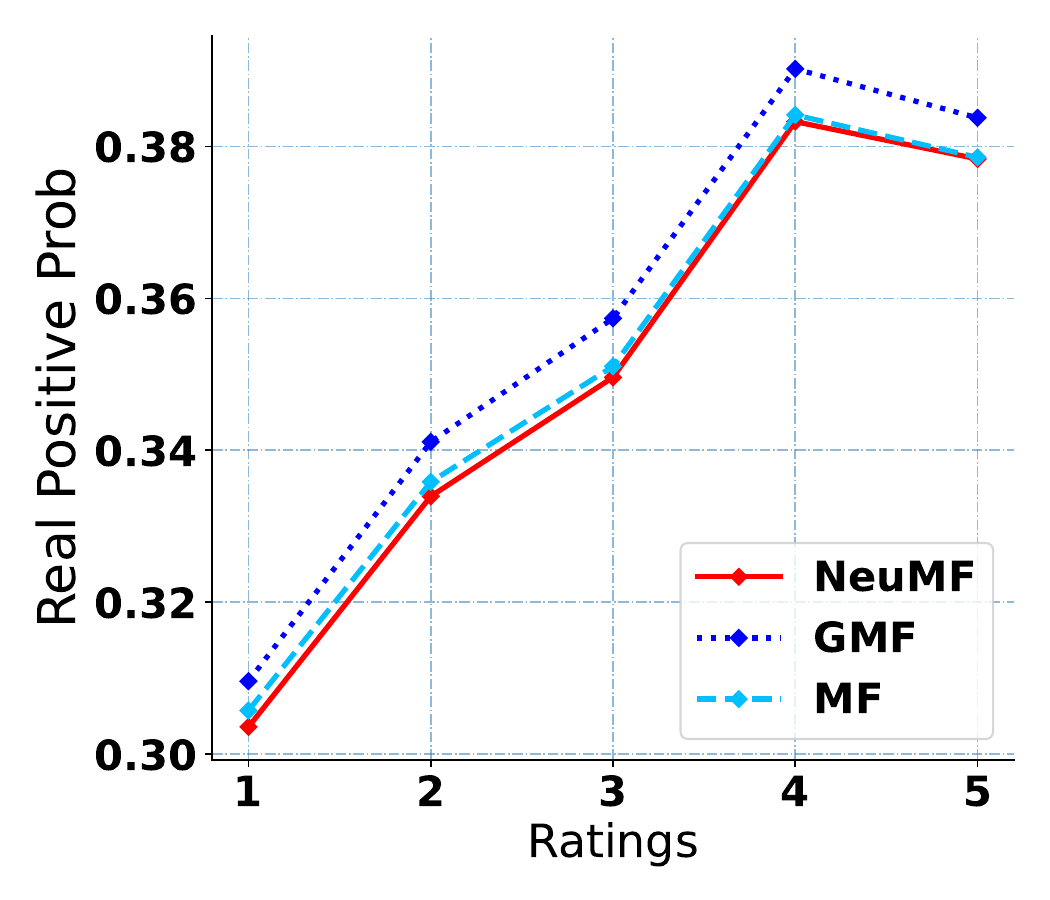}}
\subfigure[Modcloth-DeCA(p)]{\label{fig:DVAE_modcloth_rating}\includegraphics[width=0.16\textwidth]{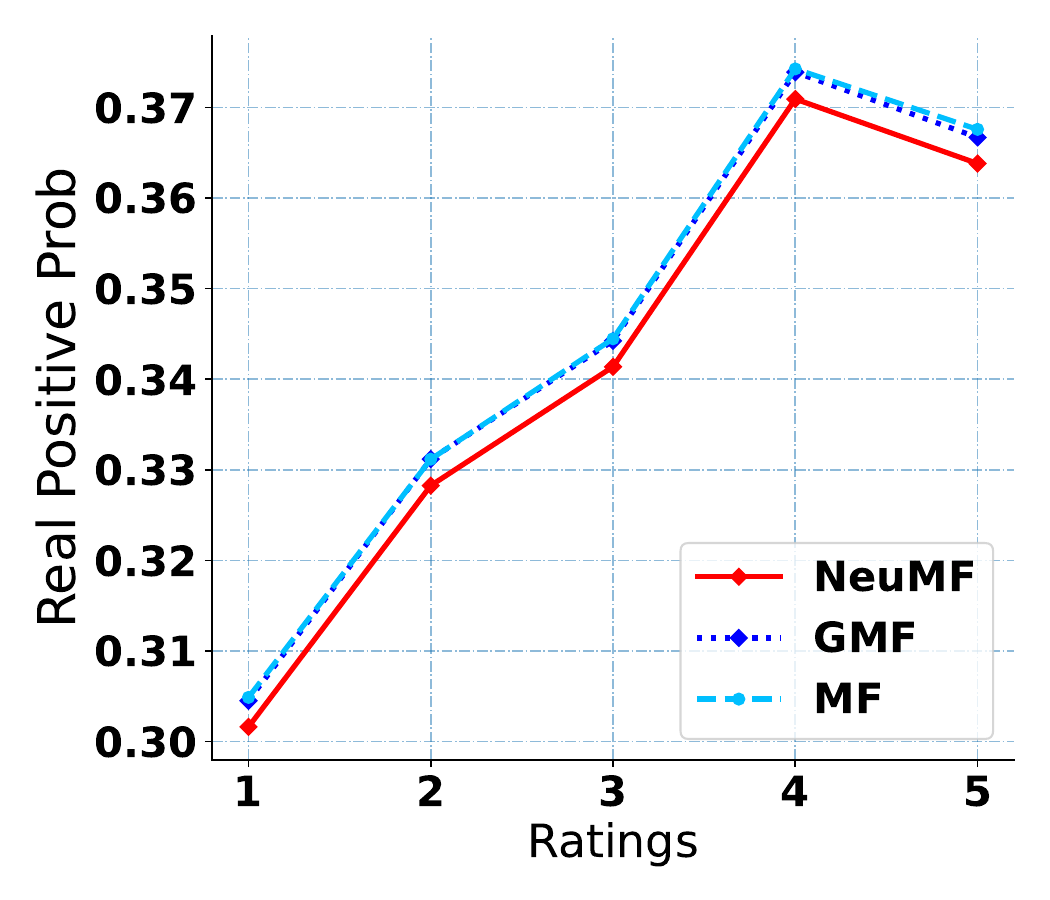}}
\subfigure[Electronics-DeCA]{\label{fig:DPI_electronics_rating}\includegraphics[width=0.16\textwidth]{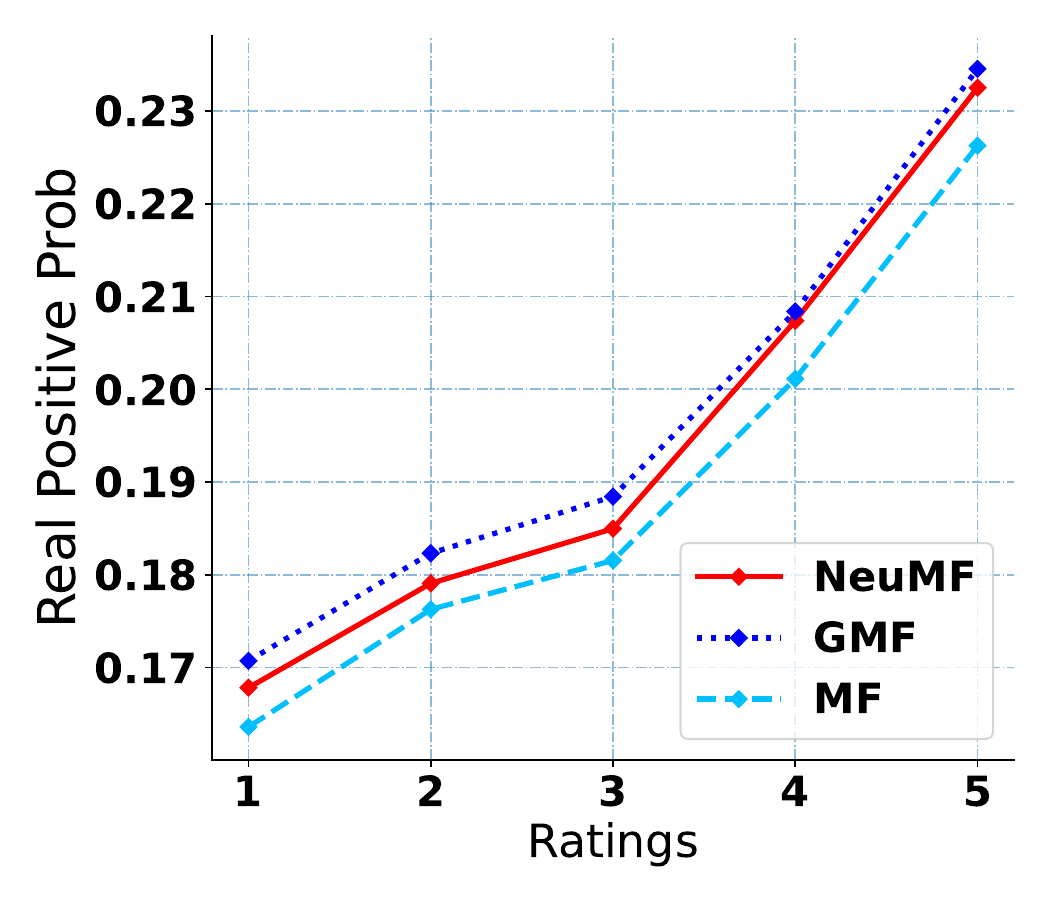}}
\subfigure[Electronics-DeCA(p)]{\label{fig:DVAE_electronics_rating}\includegraphics[width=0.16\textwidth]{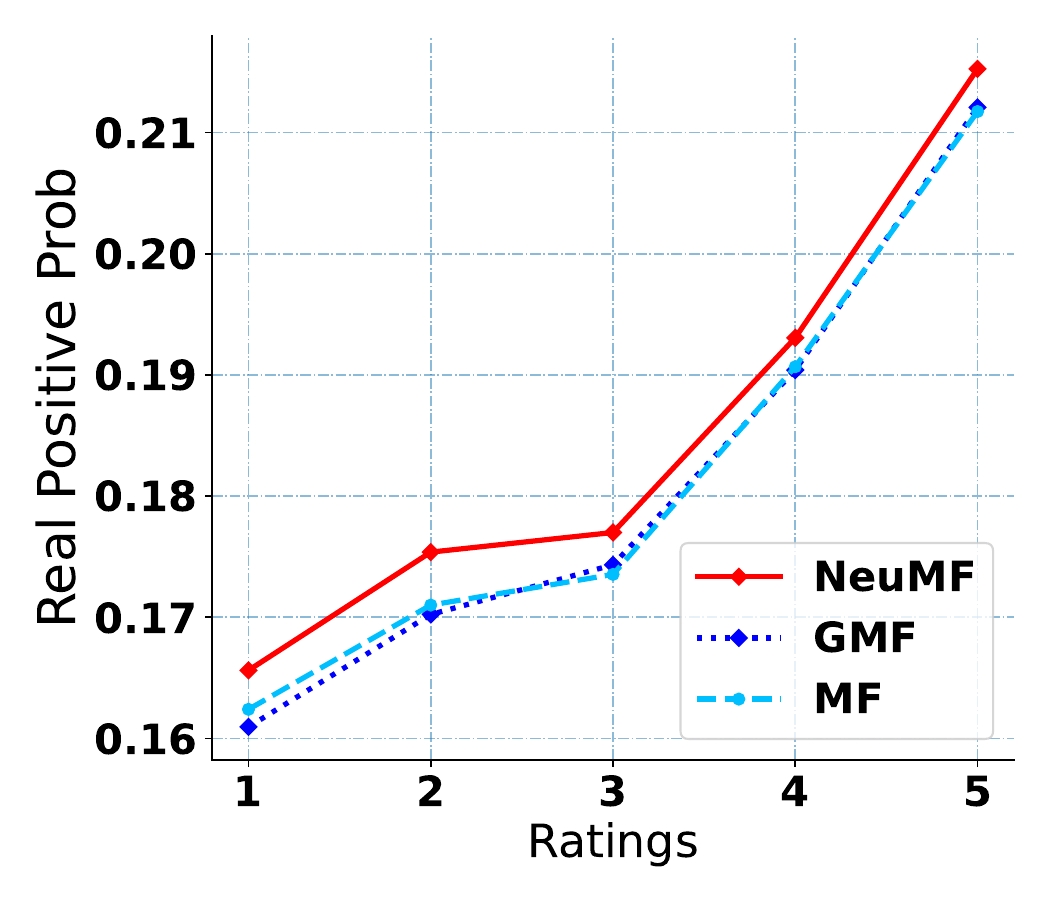}}
\vspace{-2pt}
\caption{Mean real positive probability of different ratings on three datasets.}
\label{mean_real_positive_probability}
\end{figure*}

% subsubsection Effect_of_Model_Selection (end)
\vspace{-5pt}
\subsection{Method Interpretation (RQ3)} % (fold)
\label{ssub:method_interpretation}
In this part, we conduct experiments to see whether DeCA and DeCA(p) generate the reasonable user preference distribution given the corrupted data, which can be used to show the interpretability of our methods.
For DeCA, we have the following equation:
\begin{align*}
  P(r_{ui}&=1|\tilde{r}_{ui}=1) =\frac{P(\tilde{r}_{ui}=1|r_{ui}=1)P(r_{ui}=1)}{P(\tilde{r}_{ui}=1)} 
  \tag{\stepcounter{equation}\theequation} \\
  =&\frac{h'_\psi(u,i) f_\theta(u, i)}{h'_\psi(u,i) f_\theta(u, i) + h_\phi(u,i)(1-f_\theta(u,i))}
\end{align*}
For DeCA(p), $P(r_{ui}=1|\tilde{r}_{ui}=1)$ can be directly computed as $P(r_{ui}=1|\tilde{r}_{ui}=1)= f_\theta(u, i)$ with condition $\tilde{r}_{ui}=1$.
$P(r_{ui}=1|\tilde{r}_{ui}=1)$ describes the probability of an interacted sample to be real positive. Figure \ref{mean_real_positive_probability} shows the relationship between ratings and this mean real positive probability on the datasets MovieLens, Modcloth and Electronics. 
We can see that the probability gets larger as ratings go higher, which is consistent with the impression that examples with higher ratings should be more likely to be real positive. 
%indicate that our methods are capable of detecting noisy samples, thus to avoid the model to be affected by these noisy samples. 
Besides, we can see that the probability trend is consistent with different $h$ and $h'$ models, which further demonstrate the generalization ability of our methods.

\begin{table}
\centering
\caption{Effect of model selection on Modcloth.}
\vspace{-0.3cm}
\label{tab:GMF_modcloth_gamma}
\begin{tabular}{c|c|cccc}
\hline
Method & $h$ and $h'$ & R@5 & R@20 & N@5 & N@20 \\
\hline 
\hline
\multicolumn{2}{c|}{Normal} & 0.0629 & 0.2246 &  0.0430 & 0.0884  \\
\hline
\multirow{3}{*}{DeCA}& MF&0.0717 & 0.2452 & 0.0500 & 0.0985 \\
& GMF& 0.0740 & 0.2453 & 0.0513 & 0.0989 \\
& NeuMF& 0.0747 & 0.2448 & 0.0519 & 0.0991 \\
\hline
\multirow{3}{*}{DeCA(p)}& MF& 0.0743 & 0.2465 & 0.0515 & 0.0996 \\
& GMF& 0.0740 & 0.2464 & 0.0508 & 0.0989 \\
& NeuMF& 0.0751 & 0.2464 & 0.0514 & 0.0992 \\
\hline
\end{tabular}
\vspace{-10pt}
\end{table}

%% file: 5.related_work.tex
\section{Related Work} % (fold)
\label{sec:related_work}
Modern recommender systems are usually trained using implicit feedback.
The training of implicit recommenders usually needs to sample negative examples from missing interactions and then feed the interacted items and sampled negative instances for pair-wise ranking \cite{rendle2014bayesian} or BCE loss function. Besides, there are also attempts to investigate non-sampling approaches \cite{he2016fast,chen2019efficient,yuan2018fbgd} for implicit feedback.
Regarding the recommendation models, MF \cite{koren2009matrix} is one of the most notable and effective models.%, which projects users and items to embedding vectors and then calculate the inner product between them as the prediction score.
Recently, plenty of work has proposed deep learning-based recommendation models, such as GMF, NeuMF\cite{he2017neural}, CDAE\cite{wu2016collaborative} and Wide\&Deep \cite{he2017neural}. The key idea is to use deep learning to increase model expressiveness. Besides, graph neural networks also demonstrated their capability in recommendation.
Plenty of models have emerged, such as HOP-Rec \cite{yang2018hop}, KGAT \cite{wang2019kgat}, NGCF \cite{wang2019neural} and LightGCN \cite{he2020lightgcn}.

Recently, there has been some researches \cite{jagerman2019model,lu2018between} pointing out that the observed implicit feedback could be easily corrupted by different factors, such as popularity bias, exposure bias and so on \cite{chen2020bias}. 
As a result, there have been some efforts aiming to address the noisy problem, which can be categorized into re-sampling methods \cite{yu2020sampler,gantner2012wbpr,ding2019samplerview,ding2018improvedview,ding2019reinforced}, re-weighting methods \cite{wang2020denoising,shen2019learning, nguyen2019self} and methods utilizing additional knowledge input \cite{kim2014modeling,zhao2016gaze,lu2019effects}.
Re-sampling methods aim to design more effective samplers. For example, \cite{gantner2012wbpr} proposed to sample popular but not interacted items as negative examples while \cite{ding2018improvedview} proposed that the viewed but not purchased items are highly likely to be real negative. \cite{ding2019reinforced,wang2020reinforced} proposed to use reinforcement learning for negative sampling. The performance of re-sampling depends heavily on the sampling strategy \cite{yuan2018fbgd}, which is usually developed heuristically \cite{yu2020sampler}. Re-weighting methods usually identify the noisy examples as samples with high loss values and then assign lower weights to them. 
Besides, additional knowledge such as dwell time\cite{kim2014modeling}, gaze pattern\cite{zhao2016gaze} and auxiliary item features\cite{lu2019effects} can also be used to denoise implicit feedback.

%% file: 6.conclusion.tex
\vspace{3pt}
\section{Conclusion} % (fold)
\label{sec:conclusion}
In this work, we propose model-agnostic training frameworks to learn robust recommenders from implicit feedback. We find that different models tend to make more consistent agreement predictions for clean examples compared with noisy ones.
To this end, we propose \emph{denoising with cross-model agreement} (DeCA), which utilizes predictions 
from different models as the denoising signal. We employ the proposed methods on four recommendation models and conduct extensive experiments on four datasets. The results demonstrate the effectiveness of our methods.  
To the best of our knowledge, this is the first paper to consider the agreement across different models as the denoising signal, and it's also the first work to denoise both noisy positive examples and noisy negative examples. Future work includes using DeCA for other domain applications and generalizing DeCA for multi-class scenarios.